\documentclass[final,5p,times,twocolumn,number]{elsarticle}

\usepackage{amssymb}
\usepackage{amsmath}
\usepackage{color}
\newtheorem{lemma}{Lemma}

\usepackage{mhchem}
\usepackage{booktabs,rotating}
\usepackage{multirow}

\journal{XXX}

\begin{document}

\begin{frontmatter}



\title{Optimal Strategy for Stabilizing Protein Folding Intermediates}


\author[first]{Mengshou Wang}
\ead{wangmsh6@mail2.sysu.edu.cn}

\author[second]{Liangrong Peng\corref{cor1}}
\ead{peng@mju.edu.cn}

\author[third]{Baoguo Jia}
\ead{mcsjbg@mail.sysu.edu.cn}

\author[first]{Liu Hong\corref{cor1}}
\cortext[cor1]{Corresponding author}
\ead{hongliu@sysu.edu.cn}

\affiliation[first]{organization={School of Mathematics, Sun Yat-sen University},
            addressline={135 Xingang West Road},
            city={Guangzhou},
            postcode={510275},
            state={Guangdong Province},
            country={P.R.C.}}

\affiliation[second]{organization={College of Mathematics and Data Science, Minjiang University},
	addressline={200 Xiyuangong Road, Shangjie Town, Minhou County},
	city={Fuzhou},
	postcode={350108},
	state={Fujian Province},
	country={P.R.C.}}

\affiliation[third]{organization={School of Science, Sun Yat-Sen University},
	addressline={No. 66 Gongchang Road, Guangming District},
	city={Shenzhen},
	postcode={518107},
	state={Guangdong Province},
	country={P.R.C.}}

\begin{abstract}

To manipulate the protein population at certain functional state through chemical stabilizers is crucial for protein-related studies. It not only plays a key role in protein structure analysis and protein folding kinetics, but also affects protein functionality to a large extent and thus has wide applications in  medicine, food industry, etc.  However, due to concerns about side effects or financial costs of stabilizers, identifying optimal strategies for enhancing protein stability with a minimal amount of stabilizers is of great importance. Here we prove that either for the fixed terminal time (including both finite and infinite cases) or the free one, the optimal control strategy for stabilizing the folding intermediates with a linear strategy for stabilizer addition belongs to the class of Bang-Bang controls. The corresponding optimal switching time is derived analytically, whose phase diagram with respect to several key parameters is explored in detail. The Bang-Bang control will be broken when nonlinear strategies for stabilizer addition are adopted. Our current study on optimal strategies for protein stabilizers not only offers deep insights into the general picture of protein folding kinetics, but also provides valuable theoretical guidance on treatments for protein-related diseases in medicine.
\end{abstract}




\end{frontmatter}




\section{Introduction}
\label{introduction}

Proteins are the primary executors of various physiological functions within living organisms, such as enzymes with catalytic functions, insulin that regulates metabolism, and immune proteins involved in the body's defense mechanisms \cite{dill2012protein, goldberg1995three, knighton1991crystal, proud2006regulation,di2023control, kaufmann1990heat}. The functionality of proteins is closely associated with their three-dimensional structures. Understanding how a linear polypeptide chain forms a complex three-dimensional structure is a critical issue in the field of molecular biology and known as the protein folding problem \cite{anfinsen1961kinetics}. Recently, the emergence of deep-learning-based approaches, such as AlphaFold \cite{senior2020improved}, AlphaFold2 \cite{jumper2021highly} and RoseTTAFold \cite{baek2021accurate}, capable of accurately predicting the three-dimensional structures of nearly all proteins based on their primary sequences, has made a significant breakthrough towards the problem of protein structure prediction. However, the detailed dynamics of how proteins transit from unfolded states to the folded state remains unclear \cite{lutomski2022next}.

According to Christian Anfinsen, the native structure of a protein is its lowest free energy state under normal folding conditions \cite{anfinsen1961kinetics}. Then the real protein folding pathway follows the one with the largest probability among all possible trajctories linking the unfolded state to the native state on the free energy landscape \cite{Dill1997, wolynes1995navigating,Bryngelson1995-kf,onuchic1997theory}. However, with an increase in the length of protein chains, the free energy landscape will become more and more rugged \cite{hartl2009converging,jahn2008folding}, leading to a exponentially explosion in the number of intermediate states and possible folding pathways. As a consequence, finding out the correct protein folding pathway is NP hard in nature \cite{unger1993finding,beygmoradi2023recombinant}.  

To simplify the whole physical picture of protein folding kinetics, plenty of phenomenological models have been proposed in the past decades, among which the two-state and three-state models are the most classical ones. The former simplifies the whole protein folding procedure into a conformational transition between the unfolded and folded states; while the latter includes an additional intermediate state (or the partially folded state), which plays a significant role in the folding kinetics of large proteins \cite{zwanzig1997two,fersht1999structure}. Despite of their simplicity, the two-state and three-state models correctly grasp several key features of protein folding kinetics, and provide a straight way to explain experimental observations both in quality and in quantity.

In this study, we focus on the problem of manipulating the protein population at certain functional state, e.g. the partially folded state. This problem is crucial for plenty of protein related studies. Firstly, in the field of protein structure analysis \cite{xiao2023structure, gupta2021structure}, a key issue is how to amplify the population of proteins at certain functional state, so that one can adopt further advanced tools, like NMR, cryo-electron microscopy, etc., to perform further detailed investigations. Secondly, to control the protein population means that we are able to manipulate its functionality to a great extent. For example, many amyloid-related proteins exhibit marginal stability \cite{taverna2002proteins,goldenzweig2018principles}. The misfolding of even a small amount of these proteins can disrupt proteostasis, potentially leading to severe neurodegenerative diseases, such as Alzheimer's, Parkinson's, and Huntington's diseases \cite{knowles2014amyloid,chiti2017protein,KHAN2022143,CHAARI2013196}. 
Thirdly, the nature of many commonly used approaches during the study of protein folding kinetics, like mutations in the amino acid sequence or changes in the environment, such as pH, temperature, salt concentrations, is to modulate protein populations at different states \cite{stein2019biophysical,yue2005loss}. In this way, transitions between several interested states are highlighted and will be allowed for investigation one by one.

To start with, we refer to the classical two-state and three-state models for protein folding kinetics \cite{tanford1970protein,schindler1995extremely,jackson1998small,roder1997kinetic,fersht1999structure,nolting2005protein}. Both in-vitro and in-vivo conditions are considered here, the latter of which includes additional processes for protein generation and degradation. Then we study the optimal control problem aiming at stabilizing those proteins belonging to either the folded state for the two-state model or the partially folded state for the three-state model. To be specific, our goal is achieved through the introduction of  chemical stabilizers, which are usually small molecules and can interact with proteins adopted specific conformations (or at the desired functional state). For example, the effects of several chemicals, including mannitol, betaine, trehalose, taurine, linoleic acid, beta cyclodextrin, copper sulfate, spermidine, maltose, maltodextrin, sucrose, dextran, beta alanine, myo-inositol and cysteine, on erythropoietin stabilization were monitored \cite{Mortazavi2018physico}. Then, the optimal control theory is utilized to provide analytical solutions to the problems of fixed finite terminal time, fixed infinite terminal time, free terminal time, nonlinear additions of stabilizers, etc. The influence of key parameters on the optimal strategy is thoroughly explored too. Our discussions on optimal strategies for stabilizing protein folding intermediates shed new light on protein folding kinetics, and provide theoretical guidance for treating protein-related diseases, like Alzheimer's, in medicine.

\section{Methods}

\subsection{Kinetic models for protein folding faciliated by stabilizers}
\label{subsec2.1}
\begin{figure}
	\centering
	\includegraphics[width=0.5\textwidth, angle=0]{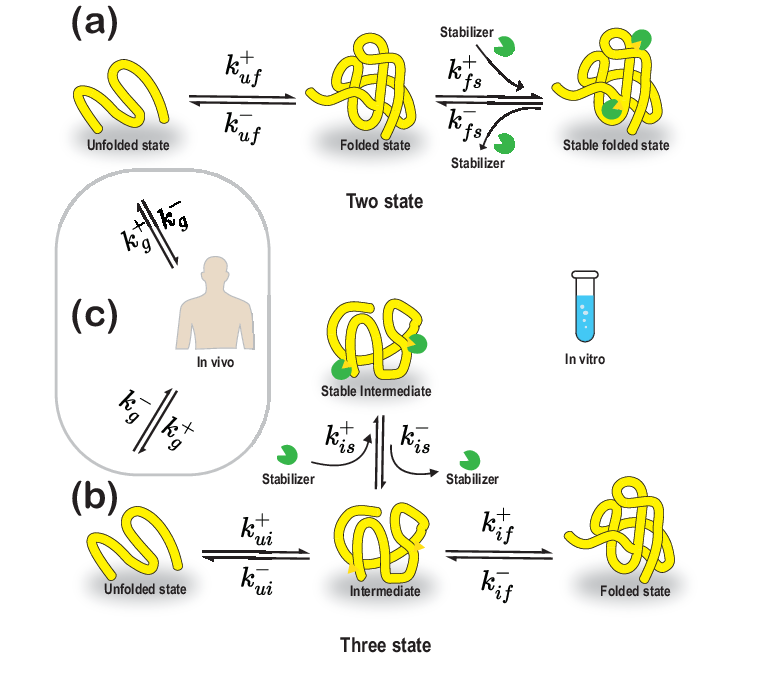}	
	\caption{Schematic diagram for the two-state and three-state protein folding processes. Interactions between stabilizers and proteins at the folded state for the two-state model or at the intermediate state for the three-state model are considered too. (a-b) illustrate the two-state model and the three-state model in vitro respectively. (c) depicts the in-vivo situations by including protein generation and degradation.}
	\label{fig_1}%
\end{figure}

To fulfill their diverse biofunctions, proteins usually enjoy highly flexible structures and can freely transit between a number of different conformations. Thereby, we adopt the classical two-state and three-state models to characterize the dynamic evolution of protein strctures. The in-vitro and in-vivo situations are considered separately. Their difference relies on the presence of processes for protein generation and degradation or not. For the purpose of amplifying the population of proteins at certain desired state, chemical stabilizers are introducted into the system, which have strong binding affinity with the proteins. Without loss of generality, here we assume the stabilizers can only interact with proteins belonging to either the folded state for the two-state model or the partilly folded state for the three-state model, as illustrated in Fig. \ref{fig_1}.

In vitro, the two-state model for protein folding faciliated by chemical stabilizers can be represented by the following chemical reactions:
\begin{equation}\label{in vitro two-state}
	\ce{U <=>[k^{+}_{uf}][k^{-}_{uf}] F},\quad \ce{F + S <=>[k^{+}_{fs}][k^{-}_{fs}] FS},
\end{equation}
where $\mathrm{U}$, $\mathrm{F}$, and $\mathrm{FS}$ denote respectively the unfolded, folded, and stabilized folded proteins, $\mathrm{S}$ denotes the stabilizer. $k^{\pm}_{uf}$ (resp. $k^{\pm}_{fs}$) are the forward/backward reaction rate constants for protein folding (resp. binding and unbinding a folded protein by the stabilizer). 
The uppercase letter $K_{x} = k^{+}_{x}/k^{-}_{x}$ is defined for the equilibrium constant, e.g. $K_{uf} = k^{+}_{uf}/k^{-}_{uf}$.

Similarly, the chemical reactions for the three-state model are represented as
\begin{equation}\label{in vitro three-state}
	\ce{U <=>[k^{+}_{ui}][k^{-}_{ui}] I <=>[k^{+}_{if}][k^{-}_{if}] F},\quad \ce{I + S <=>[k^{+}_{is}][k^{-}_{is}] IS},
\end{equation}
where $\mathrm{I}$ and $\mathrm{IS}$ denote the partially folded and stabilized partially folded proteins. $k^{\pm}_{ui}$, $k^{\pm}_{if}$ and $k^{\pm}_{is}$ are the reaction rate constants for conversions among the unfolded, partially folded and folded states, as well as the binding and unbinding of stabilizers with partially folded proteins, respectively.

Compared to the in-vitro scenario, additional processes involving the generation and degradation of unfolded proteins are introduced in vivo (see Fig. \ref{fig_1}(c)). These new reactions are given by
\begin{equation}\label{in vivo}
	\ce{$\emptyset$ <=>[k^{+}_{g}][k^{-}_{g}] U},
\end{equation}	
where $k^{+}_{g}$ and $k^{-}_{g}$ denote the rate constants for the generation and degradation of unfolded proteins.

With respect to the chemical reactions mentioned above, we can formulate the kinetic equations for both the two-state model as
\begin{equation}\label{two-state}
	\begin{aligned}
		&\dot{[\mathrm{U}]}(t) = -k^{+}_{uf}[\mathrm{U}]+k^{-}_{uf}[\mathrm{F}] \underline{-k^{-}_{g}[\mathrm{U}] +k^{+}_{g}},\\
		&\dot{[\mathrm{F}]}(t) = k^{+}_{uf}[\mathrm{U}]-k^{-}_{uf}[\mathrm{F}]-k^{+}_{fs}[\mathrm{F}][\mathrm{S}]+k^{-}_{fs}[\mathrm{FS}],\\
		&\dot{[\mathrm{FS}]}(t) = k^{+}_{fs}[\mathrm{F}][\mathrm{S}]-k^{-}_{fs}[\mathrm{FS}],\\
		&\dot{[\mathrm{S}]}(t) = -k^{+}_{fs}[\mathrm{F}][\mathrm{S}]+k^{-}_{fs}[\mathrm{FS}]+u,
	\end{aligned}
\end{equation}
and the three-state model as
\begin{equation}\label{three-state}
	\begin{aligned}
		&\dot{[\mathrm{U}]}(t)  = -k^{+}_{ui}[\mathrm{U}]+k^{-}_{ui}[\mathrm{I}] \underline{-k^{-}_{g}[\mathrm{U}] +k^{+}_{g}},\\
		&\dot{[\mathrm{I}]}(t) =k^{+}_{ui}[\mathrm{U}]-k^{-}_{ui}[\mathrm{I}] - k^{+}_{if}[\mathrm{I}]+k^{-}_{if}[\mathrm{F}]-k^{+}_{is}[\mathrm{I}][\mathrm{S}]+k^{-}_{is}[\mathrm{IS}],\\
		&\dot{[\mathrm{F}]}(t)  = k^{+}_{if}[\mathrm{I}]-k^{-}_{if}[\mathrm{F}],\\
		&\dot{[\mathrm{IS}]}(t)  = k^{+}_{is}[\mathrm{I}][\mathrm{S}]-k^{-}_{is}[\mathrm{IS}],\\
		&\dot{[\mathrm{S}]}(t)  = -k^{+}_{is}[\mathrm{I}][\mathrm{S}]+k^{-}_{is}[\mathrm{IS}]+u.
	\end{aligned}
\end{equation}
where the symbol ``[]'' denotes the molecular concentration of the corresponding chemical species (e.g. $[\mathrm{U}]$ represents the molecular concentration of unfolded proteins). The underscored portion indicates terms for the in-vivo scenario, while $u$ represents the addition rate of stabilizers.

\subsection{Steady-state analysis}
\label{subsec2.2}
To achieve the goal of manipulating the amount of folded or partial folded proteins, especially at the steady state, we perform mathematical analyses on the long-term behaviors of protein folding kinetics and reveal the optimal dosage of stabilizers. It is straightforward to show that the steady state of the system solely depends on the total addition amount of stabilizers, irrespective of how they are added, when the degradation of stablizers is omitted.

Results for the steady-state analyses on the four models established in Section \ref{subsec2.1}, namely the in-vitro two-state model (case 1), in-vitro three-state model (case 2), in-vivo two-state model (case 3), and in-vivo three-state model (case 4), are summarized in Table \ref{Table1}. Details of derivation can be found in \ref{Appendix A1} to \ref{Appendix A3}.

\begin{sidewaystable*}
	\centering
	\begin{tabular}{@{}clllc@{}}
		\toprule
		\multicolumn{2}{c}{Model}              & \multicolumn{1}{c}{Without Stabilizer} & \multicolumn{1}{c}{With Stabilizer}&\multicolumn{1}{c}{Stability (With Stabilizer)} \\ \midrule
		\multirow{7}{*}{In-vitro} & Two-state                          &$\begin{aligned}
			&[\mathrm{U}]^{ss}_0 = P_{total}/(K_{uf}+1),\\ \cmidrule(l){2-4}
			&[\mathrm{F}]^{ss}_0 = P_{total}K_{uf}/(K_{uf}+1),
		\end{aligned}$    & $
		\begin{aligned}
			&[\mathrm{U}]^{ss} = \frac{1}{2}(\sqrt{\eta_{1}^2+4\kappa_{1} P_{total}}-\eta_{1})/(1+K_{uf}),\\
			&[\mathrm{F}]^{ss} = K_{uf}[\mathrm{U}]^{ss},\\
			&[\mathrm{FS}]^{ss}  = P_{total} -[\mathrm{U}]^{ss} -[\mathrm{F}]^{ss},\\
			&[\mathrm{S}]^{ss}  = S_{total}-P_{total}+[\mathrm{U}]^{ss}+ [\mathrm{F}]^{ss},
		\end{aligned}
		$& Locally asymptotically stable                      \\ \cmidrule(l){2-5}
		& Three-state                         &      $\begin{aligned}
			&[\mathrm{U}]^{ss}_0 = P_{total}/(1+K_{ui}+K_{ui}K_{if}),\\
			&[\mathrm{I}]^{ss}_0 = P_{total}K_{ui}/(1+K_{ui}+K_{ui}K_{if}),\\
			&[\mathrm{F}]^{ss}_0 = P_{total}K_{ui}K_{if}/(1+K_{ui}+K_{ui}K_{if}),
		\end{aligned}$  &$
		\begin{aligned}
			&[\mathrm{U}]^{ss} = \frac{1}{2}(\sqrt{\eta_{2}^2+4\kappa_{2} P_{total}}-\eta_{2})/(1+K_{ui}+K_{ui}K_{if}),\\
			&[\mathrm{I}]^{ss} = K_{ui}[\mathrm{U}]^{ss},\\
			&[\mathrm{F}]^{ss} = K_{ui}K_{if}[\mathrm{U}]^{ss},\\
			&[\mathrm{IS}]^{ss}  = P_{total} - [\mathrm{U}]^{ss} - [\mathrm{I}]^{ss} - [\mathrm{F}]^{ss},\\
			&[\mathrm{S}]^{ss}  = S_{total}-P_{total}+[\mathrm{U}]^{ss}+[\mathrm{I}]^{ss}+ [\mathrm{F}]^{ss},
		\end{aligned}
		$&    Locally asymptotically stable                        \\ \midrule
		\multirow{6}{*}{In-vivo}  & Two-state  &$
		\begin{aligned}
			&[\mathrm{U}]^{ss}_0 = K_{g},\\
			&[\mathrm{F}]^{ss}_0 = K_{g}K_{uf},
		\end{aligned}
		$       & $\begin{aligned}
			&[\mathrm{U}]^{ss} = K_{g},\\
			&[\mathrm{F}]^{ss} = K_{uf}K_{g},\\
			&[\mathrm{S}]^{ss} = S_{total}/(1+K_{fs}K_{uf}K_{g}),\\
			&[\mathrm{FS}]^{ss} = S_{total}K_{fs}K_{uf}K_{g}/(1+K_{fs}K_{uf}K_{g}).
		\end{aligned}$                             & *                 \\ \cmidrule(l){2-5}
		& Three-state&$
		\begin{aligned}
			&[\mathrm{U}]^{ss}_0 = K_{g},\\
			&[\mathrm{I}]^{ss}_0 = K_{g}K_{ui},\\
			&[\mathrm{F}]^{ss}_0 = K_{g}K_{ui}K_{if},
		\end{aligned}
		$                              &$\begin{aligned}
			&[\mathrm{U}]^{ss} = K_{g},\\
			&[\mathrm{I}]^{ss} = K_{ui}K_{g},\\
			&[\mathrm{F}]^{ss} = K_{if}K_{ui}K_{g},\\
			&[\mathrm{S}]^{ss} = S_{total}/(1+K_{is}K_{ui}K_{g}),\\
			&[\mathrm{IS}]^{ss} = S_{total}K_{is}K_{ui}K_{g}/(1+K_{is}K_{ui}K_{g}).
		\end{aligned}$ &Locally asymptotically stable  \\  \bottomrule
	\end{tabular}
	\caption{Steady states of the two-state/three-state models under both in-vitro and in-vivo conditions. The influence of stabilizers are considered too. Here $\kappa_{1} = (1+K_{uf})/(K_{uf}K_{fs})$, $\eta_{1} = S_{total}-P_{total}+\kappa_{1}$, $\kappa_{2}= (1+K_{ui}+K_{ui}K_{if})/(K_{ui}K_{fs})$ and $\eta_{2} = S_{total}-P_{total}+\kappa_{2}$, in which $P_{total}$ and $S_{total}$ represent the total amounts of proteins and stabilizers at the steady state respectively. It is noteworthy that when $ K_{if} = 0$, the results of the three-state model reduces to those of the two-state model; and when $S_{total} = 0$, the influence of stabilizers can be neglected. }
	\label{Table1}
\end{sidewaystable*}
\begin{table*}
	\centering
	\begin{tabular}{@{}cccc@{}}
		\toprule
		\multicolumn{1}{c}{}  &   Model    &   Folds      & \multicolumn{1}{c}{Stabilizer ( $S_{expected} $ )}  \\ \midrule
		\multirow{2}{*}{In-vitro} & Two-state        &      $\gamma_2$               &$ (\kappa_{1}K_{uf}\gamma_{2})/(1-K_{uf}\gamma_{2})+P_{total}K_{uf}\gamma_{2}, $ \\ \cmidrule(l){2-3}
		& Three-state       &      $\gamma_3$                      &      $ (\kappa_{2}\widetilde{K_{uf}}\gamma_{3})/(1-\widetilde{K_{uf}}\gamma_{3})+P_{total}\widetilde{K_{uf}}\gamma_{3},$ \\ \midrule
		\multirow{2}{*}{In-vivo}  & Two-state    &      $\gamma_2$  &$(K_{uf}K_{g}+1/K_{fs})\gamma_{2},$      \\ \cmidrule(l){2-3}
		& Three-state  &   $\gamma_3$  &$(K_{ui}K_{g}+1/K_{is})\gamma_{3},$              \\ \bottomrule
	\end{tabular}
	\caption{A summary on the total amount of stabilizers to be added ($S_{expected}$), in order to achieve the desired increase (indicated by folds) in the concentration of folded proteins in two-state models (or partially folded proteins in three-state models). Here $\widetilde{K_{uf}}=K_{ui}/(1+K_{ui}K_{if})$.}
	\label{Table2}
\end{table*}

\subsection{Optimal strategy for stabilizing proteins at a specific state}\label{subsed2.3}

As we have claimed, how to amplify the population of proteins at certain functional state is a key issue in protein-related studies. Therefore, in this part, we are going to explore the optimal usage of external chemical stabilizers to maximize the amount of either folded proteins for the two-state model or partially folded proteins for the three-state model. Meanwhile, the total addition amount of stabilizers is required to be minimized too, either to lower the ecomomic cost or to reduce potential toxicity of chemicals \cite{loo2007chemical}. To achieve a balance between above contradicted desires, the following objective functional for the two-state model is introduced as
\begin{equation}\label{objective functional two-state}
	J[u(\cdot)] = - w_{Tol}([\mathrm{FSF}](T))+\int_{0}^{T}u(\tau)d \tau.
\end{equation}
Here, $\mathrm{FSF}$ represents the total folded proteins, both bound and unbound by stabilizers, whose concentration thus is given by $[\mathrm{FSF}] = [\mathrm{FS}]+[\mathrm{F}]$. Additionally, T denotes the fixed finite terminal time considered in the analysis. The function $u(t)$ represents the rate of stabilizer addition over time, and  $\int_{0}^{T} u(\tau) d\tau$ representing the total amount of stabilizers added. $w_{Tol}$ is the weight between the two, representing the tolerance of stabilizers, which is typically inversely related to its side effects or financial burden, hereinafter referred to as the tolerance. It is straightforward to see that a higher $ w_{\text{Tol}} $ warrants a greater allowance for the stabilizer's usage $\int_{0}^{T}u(\tau)d \tau$.

Similarly, for the three-state model, we can establish the objective functional,
\begin{equation}\label{objective functional three-state}
	J[u(\cdot)] = - w_{Tol}([\mathrm{ISI}](T))+\int_{0}^{T}u(\tau)d \tau.
\end{equation}
Here $[\mathrm{FSF}]$ is replaced by $[\mathrm{ISI}]= [\mathrm{IS}]+[\mathrm{I}]$, the latter of which represents the total concentration of partially folded proteins.

To be specific, we impose a pratical constraint on the adding rate $u$, i.e.
\begin{equation}\label{constraints}
	0 \leq u(t) < u_{\max},\quad \forall  t \in [0, T],
\end{equation}
where $ u_{\max} $ represents the maximal rate for stabilizer addition.

With respect to above elements, the optimal control problem is established accordingly, whose solution could be derived based on the Pontryagin's maximum principle. All details are summarized in \ref{Appendix C}.

\subsection{Software and library}
The numerical calculations in this paper is performed by MATLAB (version R2023a), with relevant libraries including the Optimization Toolbox (version 9.5) and OpenOCL \cite{koenemann2017openocl}.

\section{Results}
In this section, we will focus on the control of the steady state and its efficiency as discussed before. In particular, we will make a thorough exploration on the optimal control strategy for the fixed terminal time, the correspinding approximate optimal switch time (OST) as well as  phase diagrams highlighted the influence of key parameters on the OST.
\subsection{Control of steady state}
We focus on the control of protein stability in this part, that is, how to use stabilizers to achieve the desired steady state of folded or partially folded proteins, and how much amount of stabilizers should be used in total.
In \ref{Appendix A3}, we have shown that when the degradation of stabilizers are omitted, the steady states of both folded and partially folded proteins are unrelated to the way how stabilizers are used, but rather depend solely on the total amount of stabilizers added. As a consequence, here we only consider how to design the total amount of stabilizers $S_{expected}$ to achieve the steady state of folded or partially folded proteins as expected.

For the two-state model, the increased folds of folded proteins at the steady state due to the addition of stabilizers are given by
$$\gamma_{2} = ([\mathrm{FSF}]^{ss}-[{\mathrm{FSF}}]^{ss}_0)/ [{\mathrm{FSF}}]^{ss}_0,$$
where $[{\mathrm{FSF}}]^{ss}$ and $[{\mathrm{FSF}}]^{ss}_0$ represent concentrations of folded proteins in the presence and absence of stabilizers at the steady state. Similarly, for the three-state model,  the addition of stabilizers will lead to an increase in partially folded proteins at the steady state. The increased folds read
$$\gamma_{3} = ([{\mathrm{ISI}}]^{ss}-[{\mathrm{ISI}}]^{ss}_0)/[{\mathrm{ISI}}]^{ss}_0,$$
where $[{\mathrm{ISI}}]^{ss}$ and $[{\mathrm{ISI}}]^{ss}_0$ represent steady-state concentrations of partially folded proteins in the presence and absence of stabilizers separately.

In order to make folded or partially folded proteins to achieve the desired increased folds ($\gamma_2$ or $\gamma_3$), results on the total amount of stabilizers needed are summarized in Table \ref{Table2} and Fig. \ref{fig_2}, while detailed analyses can be found in \ref{Appendix B}. It is observed that upon introducing the parameter \(\widetilde{K_{uf}} = K_{ui}/(1+K_{ui}K_{if})\), \(S_{expected}\) takes the same form for both two-state and three-state models in vitro. Especially, when \(K_{uf}\ll1\) and \(K_{uf}K_{fs}P_{total}\gg1\) for the two-state model (\(\widetilde{K_{uf}}\ll1\) and \(\widetilde{K_{uf}}K_{fs}P_{total}\gg1\) for the three-state model), we have \(S_{expected} \propto  \gamma_{2}\) (or \(S_{expected} \propto  \gamma_{3}\)). A similar conclusion could be reached for the two-state and three-state models in vivo, as illustrated in Fig. \ref{fig_2}(a). Fig. \ref{fig_2}(c) further shows that $[\mathrm{FSF}]$ or $[\mathrm{ISI}]$ ultimately reach their expected values under four scenarios respectively, thereby confirming our theoretical findings through a numerical way.

\subsection{Efficient strategy for reaching the desired steady state}
\begin{figure*}
	\centering
	\includegraphics[width=1.1\textwidth, angle=90]{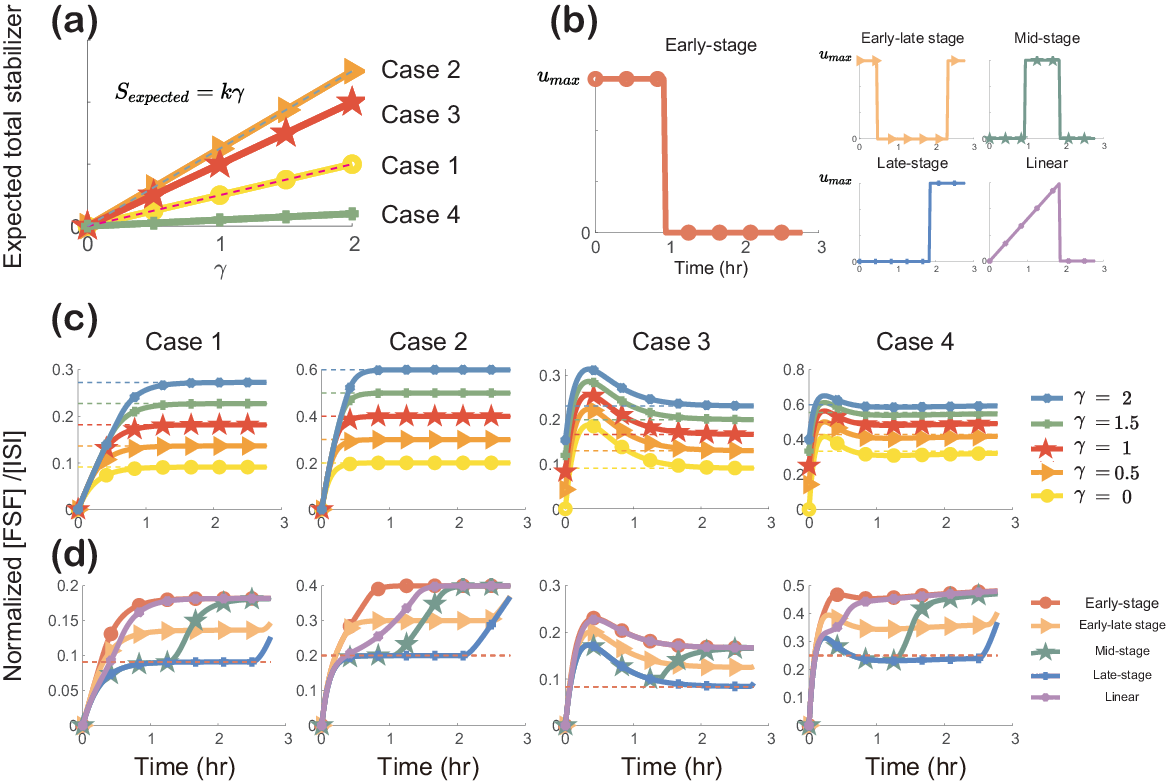}	
	\caption{Efficient strategies for stabilizer addition in four different scenarios: two states in vitro  (case 1), three states in vitro (case 2), two states in vivo (case 3), and three states in vivo (case 4). (a) demonstrates the linear or approximately linear relationship between the total amount of stabilizers $S_{expected}$ and the increased folds $\gamma_2$ or $\gamma_3$ (abbreviated as $\gamma$). (b) illustrates five typical schemes for stabilizer addition under the constraint $u(t)\leq u_{\max}$, namely the Early-stage, Early-Late stage, Mid-stage, Late-stage, and Linear usage. (c) depicts the concentration evolution of folded or partially folded proteins when stabilizers are added in the early stage. The expected increased folds are $\gamma = 0, 0.5, 1, 1.5, 2$. (d) compares the efficiency of different stabilizer addition schemes, among which the Early-stage scheme exhibits the highest efficiency among all cases.}
	\label{fig_2}%
\end{figure*}

Recall that the steady states of folded proteins or partially folded proteins depend on the total amount of stabilizers added rather than the details of usage. However, various ways of usage may influence the time when the steady state is reached, which is referred as the efficiency. In this section, we numerically explore the impacts of five different ways of usage on the efficiency with a bounded maximal addition rate $u_{\max}$. These five ways are the \emph{Early-stage} (all stabilizers are added in the early stage), \emph{Early-Late stage} (half in the early stage, half in the late stage), \emph{Mid-stage} (all in the middle stage), \emph{Late-stage} (all in the late stage), and \emph{Linear usage} (the addition rate is linearly increased in the early stage), as illustrated in Fig. \ref{fig_2}(b).

It is worth noting that the Early-stage and Linear-usage methods are similar, as stabilizer usage is concentrated on the early stage, resulting in similar effects on the folded or partially folded proteins. From Fig. \ref{fig_2}(d), it is observed that the Early-stage method can reach the steady state most rapidly across all four cases, indicating its superior efficiency. On the other hand, the rest methods lead to less efficient attainment of the steady state, with the Early-Late stage and Late-stage methods requiring more time to reach the steady state. Therefore, our preliminary conclusion is that the earlier the stabilizers are used, the higher its efficiency will be.

\subsection{Optimal strategy for stabilizing proteins at a specific state}
In the \textit{Methods}, we have established the general theory of optimal control for stabilizing proteins at a given functional state within a finite time. A general conclusion is the optimal strategy takes the form of a Bang-Bang control. However, the number of switching time points and their exact positions remain unclear, which will be explored both theoretically and numerically in this part.

Taking the in-vitro two-state model as an example (see \ref{Appendix C3}), we can prove that the addition of stabilizers follows a Bang-Bang control strategy with only one optimal switching time (denoted as $OST$). Therefore, the optimal strategy is
\begin{equation}
	u(t) = \left\{
	\begin{aligned}
		&u_{\max},\quad &0\le t<OST,\\
		&0,\quad &OST\le t<T.
	\end{aligned}
	\right.
\end{equation}
The OST corresponds to the time point where the key adjoint state $\tilde{p}$ equals to 1 ($\tilde{p}$ representing the adjoint state introduced by the differential equation constraint on $[S]$ in the Hamiltonian according to the Pontryagin maximum principle, see \ref{Appendix C2}). Before the OST, the stabilizers are applied at the maximum rate $u_{\max}$; while after the OST, no new stabilizer will be supplied.

\begin{figure*}
	\centering
	\includegraphics[width=1\textwidth, angle=0]{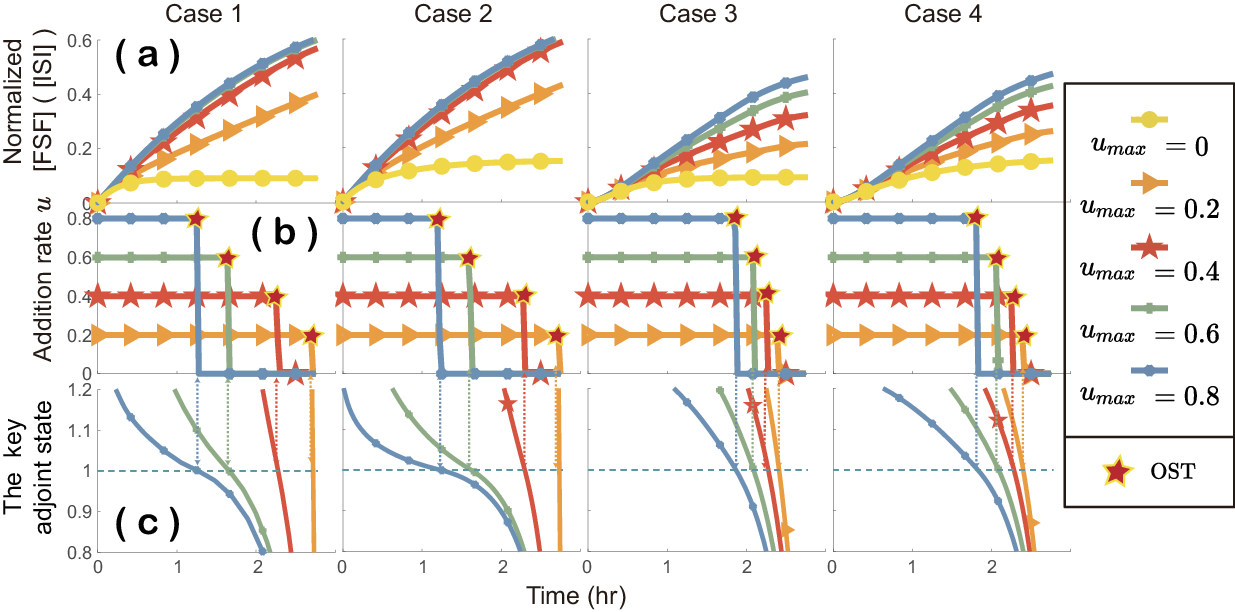}	
	\caption{The Bang-Bang control and the optimal switching time for problems with fixed finite terminal time. (a-b) Trajectories of the optimal stabilizer addition rates and the corresponding normalized $[\mathrm{FSF}]$ or $[\mathrm{ISI}]$ are illustrated for 4 cases, that is two states in vitro (case 1), three states in vitro (case 2), two states in vivo (case 3), and three states in vivo (case 4). The maximal stabilizer addition rate is set as $u_{\max} =0.2, 0.4, 0.6, 0.8$ separately. 
(c) Values of the key adjoint states can be used for determining the OST.}
	\label{fig_3}%
\end{figure*}

In Fig. \ref{fig_3}, trajectories of the optimal addition rate of stabilizers and  the corresponding normalized $\mathrm{[FSF]}$ or $\mathrm{[ISI]}$ are presented for the four cases separately. Different upper bounds of stabilizer addition rates \( u_{\max} \) have been considered too. As \( u_{\max} \) increases, more $\mathrm{FSF}$ or $\mathrm{ISI}$ are observed, and the OST is advanced as depicted in Figs. \ref{fig_3}(a)-(b). Most importantly, the correspondence between trajectories in Figs. \ref{fig_3}(b)-(c) confirms the fact that the time point where the key adjoint state equals to 1 serves exactly as the OST.

\subsection{Approximate optimal switching time}
\label{Aost}
The calculation of the optimal switching time (OST) through the key adjoint state requires to solve a complicated two-point boundary value problem. Therefore, it is essential to characterize the OST using simple algebraic expressions (see \ref{Appendix C4}). When the terminal time $T$ is sufficiently large to allow each chemical species to approximately reach their respective steady states, we can provide an approximate $OST$.

For this purpose, the effective folding index, referred as $EFI$, is introduced. The relation $w_{Tol} \le EFI$ indicates that the stabilizer has poor effectiveness and its usage is not recommended, resulting in $OST=0$. Otherwise, $w_{Tol} > EFI$ indicates that the usage of stabilizers is preferred, leading to a non-zero OST.

The following conclusions are reached (see \ref{Appendix C4}):
\begin{itemize}
	\item For the in-vitro two-state model, $EFI = 1+K_{uf}$. When $w_{Tol} > EFI$, $OST=\min\{\left[\zeta_{1}+(\psi-2)\sqrt{\kappa_{1}P_{total}/(\psi-1)}\right]/u_{\max},T\}$; especially when $w_{Tol} \gg EFI$, we have
	$OST=\min\{\left[\zeta_{1}+\sqrt{\kappa_{1}\psi P_{total}}\right]/u_{\max},T\}$;
	\item for the in-vitro three-state model, $EFI = 1+\widetilde{K_{uf}}$. When $w_{Tol} > EFI$, $OST =\min\{\left[\zeta_{2}+(\psi-2)\sqrt{\kappa_{2}P_{total}/(\psi-1)}\right]/u_{\max},T\}$; especially when $w_{Tol} \gg EFI$, we have  $OST=\min\{\left[\zeta_{2}+\sqrt{\kappa_{2}\psi P_{total}}\right]/u_{\max},T\}$;
	\item For the in-vivo two-state model, $EFI = 1+1/(K_{uf}K_{g}K_{fs})$ (resp. $EFI = 1+1/(K_{ui}K_{g}K_{is})$ for the in-vivo three-state model). In both cases, we have $OST=T$ when $w_{Tol} > EFI$.
\end{itemize}
Here $\zeta_{1} = P_{total}-\kappa_{1}$, $\zeta_{2} = P_{total}-\kappa_{2}$, $\kappa_{1} = (1+K_{uf})/(K_{uf}K_{fs})$, $\kappa_{2}= (1+K_{ui}+K_{ui}K_{if})/(K_{ui}K_{fs})$, and $\psi = w_{Tol}/EFI$.

Based on the results for the approximate OST above, it can be concluded that for the in-vivo models, the OST will increase with the extension of the terminal time and approximately equals to the terminal time. In terms of disease treatment, this implies the lifelong usage of drugs (or stabilizers). Contrarily, for the in-vitro models, as the terminal time increases, the OST approaches a finite moment.

\subsection{Phase diagram analysis}
It has been observed that both $w_{Tol}$ and $EFI$ have a significant influence on the OST. To perform a quantitative analysis, we explore the phase diagram of the OST with respect to these two parameters. 

\begin{figure*}
	\centering
	\includegraphics[width=1\textwidth, angle=0]{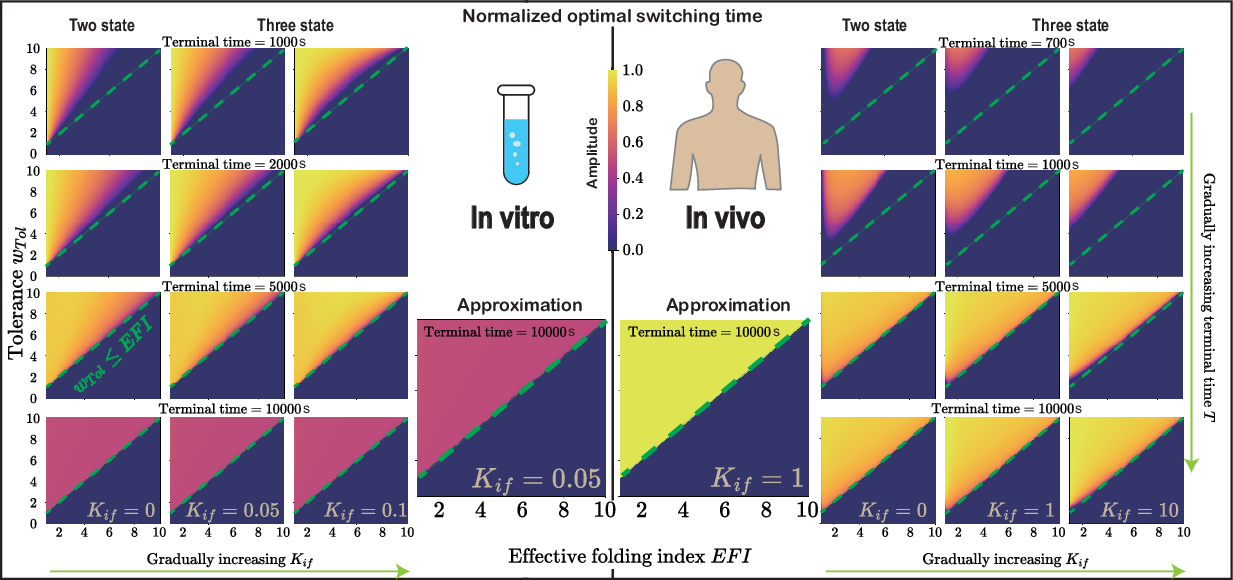}	
	\caption{Phase diagrams for the normalized optimal switching time with respect to $w_{Tol}$ and $EFI$. In the diagram, plots on the left-hand side show the  in-vitro cases, while those on the right-hand side for the in-vivo cases. Within the in-vitro cases, the two-state and three-state models are illustrated from left to right, indicated by an increasing value of $K_{if}$ as $0, 0.05, 0.1$. From top to bottom, they represent an increasing terminal time $T$. So are the in-vivo cases. Two middle panels denote results for the approximate $OST$ obtained in Sec. \ref{Aost}. The green dashed lines in all plots represent the condition $w_{Tol}=EFI$.}
	\label{fig_4}%
\end{figure*}

Consider the two-state and three-state models with different terminal times for both in-vitro and in-vivo conditions. Firstly, for all cases, when $w_{Tol}\leq EFI$ (the region below green dashed lines in Fig. \ref{fig_4}), the OST is zero, indicating no need for using stabilizers. This validates our previous findings under steady-state conditions and remains applicable for non-steady-state outcomes with smaller terminal times. Additionally, as the terminal time increases (from top to bottom in Fig. \ref{fig_4}), the boundary of the region where $OST > 0$ gradually approaches $w_{Tol} = EFI$ (indicated by the green dashed line).

Secondly, under the in-vitro scenario, as $K_{if}$ increases, the boundary of the region where $OST > 0$ approaches $w_{Tol}=EFI$; in contrast, it moves away from $w_{Tol}=EFI$ under the in-vivo scenario. When the terminal time is sufficiently large, the normalized OST gradually decreases in the region where OST$>0$ in vitro, while in vivo, it approaches 1. Our theoretical results presented in the previous section can approximate these situations well.

\section{Discussion}
Towards the key problem of mainipulating protein stability in this study, we have conducted detailed analyses from perspectives of both steady state and fixed terminal time, and have provided optimal strategies for utilizing stabilizers. Concerning the complexity of problems under consideration, our analyses presented in the maintext are far from complete. For example, the nonlinear strategies for stabilizer addition, the free terminal time as well as the inclusion of protein aggregation procedures could also play an important role. Preliminary results are presented here to inspire a more comprehensive view. 

\subsection{Nonlinear strategies for stabilizer addition}
Previously, we have assumed a linear dependence of the stabilizer concentration $[\mathrm{S}]$ on its addition rate $u$ as in Eq. \eqref{two-state}. However, when stabilizers can interact with each other and form large compounds, a sub-linear dependence would be observed. This would also happen to a system with an interface, like the lipid bilayer. How to address such nonlinear scenarios poses a big challenge. 

Here, we take the in-vitro two-state model as an illustration. A square-root relationship is considered, which leads to an alternative governing equation for $[\mathrm{S}]$ as
\begin{equation}\label{eq10}
	\dot{[\mathrm{S}]}(t) = -k^{+}_{fs}[\mathrm{F}][\mathrm{S}]+k^{-}_{fs}[\mathrm{FS}]+\sqrt{u}.
\end{equation}
The objective function and constraints remain unchanged as in the previous Sec. \ref{subsed2.3}. Correspondingly, the optimal strategy becomes a non-Bang-Bang control (see \ref{Appendix D}), i.e. 
\begin{equation}
	u(t) = \left\{
	\begin{aligned}
		&u_{\max},\quad &p_{3} \ge 2\sqrt{u_{max}},\\
		&p_{3}^2/4,\quad &0<p_{3}<2\sqrt{u_{max}},\\
		&0,\quad &p_{3} \le 0,
	\end{aligned}
	\right.
\end{equation}
where $p_{3}$ stands for the key adjoint state, which is not a constant. Therefore, $u(t)$ is not a simple piecewise constant function, and the switching points are uncertain. Similar considerations are applicable to the other models.

As shown in Fig. \ref{fig_5}(a)-(b), for the linear dependence, the addition rate of stabilizers $u(t)$ concentrates in the early stage and instantaneously becomes zero after the OST. This fact clearly states a Bang-Bang control. Contrarily, for the nonlinear dependence, the OST still exists, but the stabilizer addition rate gradually diminishes to zero after the OST instead of a sharp transition. The dramatic difference in the optimal control strategy highlights the significance of taking nonlinear effects into consideration.  

\subsection{Free terminal time}
The objective function adopted in the previous context has a fixed terminal time, and the amount of folded or partially folded proteins at the terminal time has been taken as a soft constraint. In practice, however, the terminal time is usually uncertain, and we aim at minimizing the terminal time as much as possible. Furthermore, the amount of folded (or partially folded) proteins at the terminal time will be treated as a ``hard'' constraint, meaning it needs to reach a precise quantity specified at the very beginning.

Again, we take the in-vitro two-state model for illustration. The corresponding objective function reads
\begin{equation}
	J[u(\cdot),t_{free}] = \sigma t_{free}+\int_{0}^{t_{free}}u(\tau)d \tau, 
\end{equation}
where \( t_{free} \) denotes the terminal time, which is a free variable. And $\sigma$ is the weight associated with the free terminal time. The terminal constraint for folded proteins is given by
\begin{equation}
	[\mathrm{FSF}](t_{free}) = [\mathrm{FSF}]_{tar},
\end{equation}
where \( [\mathrm{FSF}]_{tar} \) represents the targeted total amount of folded proteins. The kinetic equations and control constraints remain the same as in Sec. \ref{subsed2.3}.

In this case, the optimal control strategy becomes (see \ref{Appendix E})
\begin{equation}
	u(t) = \left\{
	\begin{aligned}
		&u_{\max},\quad &p_{3} > 1,\\
		&0,\quad &p_{3} < 1.
	\end{aligned}
	\right.
\end{equation}
The free terminal time $t_{free}$ is determined by \[H([\mathrm{U}](t),[\mathrm{F}](t),[\mathrm{S}](t),p_{1}(t),p_{2}(t),p_{3}(t),u(t))|_{t=t_{free}}=0,\] where \( H \) is the Hamiltonian in the Pontryagin maximum principle, the specific form of which can be found in \ref{Appendix D}.

In Fig. \ref{fig_5}(d), it is observed that as the weight $\sigma$ increases, the OST gradually increases, while the terminal time decreases continuously and approaches the OST. Therefore, a larger weight $\sigma$ is required in order to achieve a smaller terminal time. Two specific trajectories of the stabilizer addition rate \( u \) are given in Fig. \ref{fig_5}(c) as a comparison. 

\begin{figure}
	\centering
	\includegraphics[width=0.45\textwidth, angle=0]{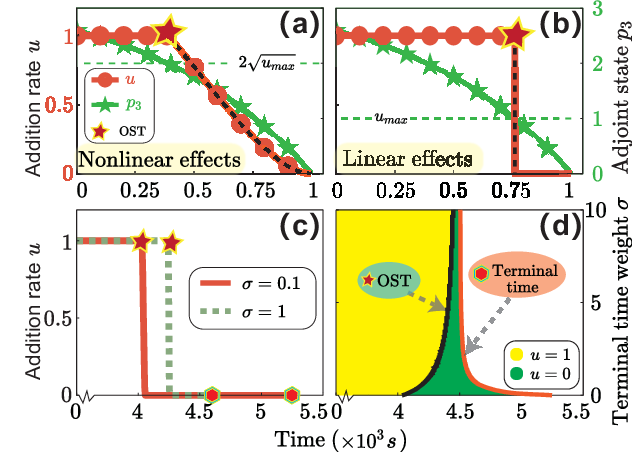}	
	\caption{Impacts of nonlinear strategie for stabilizer addition and the free terminal time. (a) and (b) depict optimal strategies for stabilizer addition with respect to the model given in Eq. \eqref{eq10} and Eq. \eqref{two-state}. The solid green line represents the key adjoint state, the dashed green lines denote 2$\sqrt{u_{max}}$ and $u_{max}$, and their intersection points give the OSTs. The black dashed lines in (a) and (b) respectively illustrate the difference between the nonlinear and linear dependence after OST -- a gradual decrease v.s. an instantaneous diminishment. (c) illustrates the optimal stabilizer addition strategies under two typical terminal time weights $\sigma = 0.1, 1$. (d) provides the phase diagram for the control problem with free terminal time. The black line denotes the OST, while the red line represents the terminal time. Stabilizers will be used only within the yellow colored region.}
	\label{fig_5}%
\end{figure}

\subsection{Inclusion of protein aggregation}
Due to the exposure of hydrophobic resiudes, protein-protein aggregation can easily happen among the partially folded or misfolded conformations. To take the influence of protein aggregation into consideration, here we refer to the sequential eglongation model for aggregation involving partially folded proteins\cite{Hong2017statis}.  
\begin{equation}\label{Cascade reaction}
	\ce{I + I <=>[k^{+}_e][k^{-}_e] I_{2}},\quad\ce{I_{2} + I <=>[k^{+}_e][k^{-}_e] I_{3}},\quad\cdots,\quad\ce{I_{n-1} + I <=>[k^{+}_e][k^{-}_e] I_{n}},
\end{equation}
where \( \mathrm{I_{j}} \) represents an aggregate containing \( j \) units of partially folded monomeric proteins \( \mathrm{I} \), for \( j = 2, \ldots, n \). Here, $n$ denotes the total number of \( \mathrm{I} \) in the largest aggregate. The rate constants for the elongation and dissociation reactions are denoted as \( k^{\pm} \). In this way, monomeric proteins are transformed into large fibrils.

Combined with the in-vitro three-state model described by the chemical reactions in \eqref{in vitro three-state}, the detailed kinetic equations for $I_j$ read 
\begin{equation}\label{three-state-cascade-reaction}
	\begin{aligned}
		\dot{[\mathrm{I}]}(t) =&k^{+}_{ui}[\mathrm{U}]-k^{-}_{ui}[\mathrm{I}] - k^{+}_{if}[\mathrm{I}]+k^{-}_{if}[\mathrm{F}]-k^{+}_{is}[\mathrm{I}][\mathrm{S}]\\
		&+k^{-}_{is}[\mathrm{IS}]+2k^{-}_e[\mathrm{I_{2}}]-2k^+_e[I]^2+\sum_{j=3}^{n}\left(k^{-}_e[\mathrm{I_{j}}]-k^{+}_e[\mathrm{I_{j-1}}][\mathrm{I}]\right),\\
		\dot{[\mathrm{I_{j}}]}(t)=&k^{+}_e([\mathrm{I_{j-1}}]-[\mathrm{I_{j}}])[\mathrm{I}]+k^{-}_e([\mathrm{I_{j+1}}]-[\mathrm{I_{j}}]),\quad j = 2,\ldots,n-1, \\
		\dot{[\mathrm{I_{n}}]}(t)=&k^{+}_e[\mathrm{I_{n-1}}][\mathrm{I}]-k^{-}_e[\mathrm{I_{n}}],
	\end{aligned}
\end{equation}
Note the extra terms appear in the second line due the inclusion of protein aggregation procedures. The equations for \({[\mathrm{U}]}\), \({[\mathrm{F}]}\), \({[\mathrm{IS}]}\), and \({[\mathrm{S}]}\) remains the same as in Eq.\eqref{three-state}.

With respect to the same objective function and constraints given in Eq.\eqref{objective functional three-state} and Eq.\eqref{constraints}, it can be shown that the corresponding optimal strategy is still a Bang-Bang control (see \ref{Appendix F}),  
However, the switching time is determined by the roots of a complex system of \( 2(4+n) \) variables $([\mathrm{U}], [\mathrm{F}], [\mathrm{IS}], [\mathrm{S}], [\mathrm{I_{j}}], j=1,\ldots,n $, and their corresponding adjoint states), which can be multiple in principle. This complexity will be explored in depth in the future.

\section{Conclusion}
In this  study, we consider both in-vitro and in-vivo conditions to investigate the optimal control problem aimed at stabilizing proteins in either the folded state for the two-state model or the partially folded state for the three-state model, by introducing a minimal amount of chemical stabilizers. We demonstrate that, both for fixed finite terminal time (including both finite and infinite cases) and for free terminal time, the optimal control strategy for stabilizing the protein folding intermediates with a linear form of stabilizer addition belongs to the class of Bang-Bang controls. The corresponding optimal switching time is derived analytically, and its phase diagram with respect to several key parameters is explored in detail. It is proven that stabilizers are valuable only under the condition \( w_{Tol} > EFI \), where $EFI$ is the effective folding index, and \( w_{Tol} \) is the tolerance of the stabilizer, which is typically inversely related to its side effects or financial cost. Our research provides theoretical guidance for the rational use of stabilizers, contributing to the appropriate medication for protein-related diseases.

\section*{CRediT authorship contribution statement}
\textbf{Mengshou Wang:} Writing – review \& editing, Writing – original draft, Visualization, Validation, Methodology, Software, Formal analysis, Conceptualization. \textbf{Liangrong Peng:} Writing – review \& editing, Visualization, Validation, Formal analysis, Funding acquisition. \textbf{Baoguo Jia:} Writing – review \& editing, Validation. \textbf{Liu Hong:} Writing – review \& editing, Visualization, Validation, Methodology, Project administration, Funding acquisition, Conceptualization.

\section*{Declaration of competing interest}
The authors declare no competing interests.

\section*{Acknowledgements}
This work was supported by the National Key R\&D Program of China
(Grant No. 2023YFC2308702), the National Natural Science Foundation of China
(12205135), Guangdong Basic and Applied Basic Research Foundation
(2023A1515010157).

\appendix

\section{Models for protein folding kinetics}
\label{Appendix A}
We will show results for protein folding kinetics under four scenarios, including the two-state and three-state models in vitro and in vivo. In the absence of stabilizers, analytical solutions will be presented; in contrast, results for the steady states and their stabilities are given by accounting for the influence of stablizers.
\subsection{Analytical solutions in the absence of stabilizers}
\label{Appendix A1}
We will first provide the explicit expressions for various chemical species during the protein folding process in the absence of stabilizers.
\paragraph{In vitro two-state model}
The chemical reactions read
\begin{equation*}\label{A1 in vitro two-state}
	\ce{U <=>[k^{+}_{uf}][k^{-}_{uf}] F},
\end{equation*}
whose corresponding kinetic equations are
$$\begin{aligned}
	\dot{[\mathrm{U}]}(t) & = -k^{+}_{uf}[\mathrm{U}]+k^{-}_{uf}[\mathrm{F}],\label{A1 in vitro two-state dyn1}\\
	\dot{[\mathrm{F}]}(t) & = k^{+}_{uf}[\mathrm{U}]-k^{-}_{uf}[\mathrm{F}]\label{A1 in vitro two-state dyn2}.
\end{aligned}$$
Due to the law of mass conservation $P_{total}=F(t)+U(t)$, the following analytical expressions can be derived, i.e.
\begin{align}
	[\mathrm{F}](t) &= C_{2}\mathrm{exp}[-(k^{+}_{uf}+k^{-}_{uf})t]+C_{1},\\
	[\mathrm{U}](t) &= P_{total} - [\mathrm{F}](t)\label{two-state sol2},
\end{align}
where $C_{1} =P_{total}K_{uf}/(K_{uf}+1), C_{2} =\mathrm{F}(0)-C_{1}$. 

Since $k^{+}_{uf}+k^{-}_{uf} > 0$,
$$
\begin{aligned}
	[\mathrm{U}](t) &\rightarrow [\mathrm{U}]_0^{ss} = P_{total}/(K_{uf}+1),\\
	[\mathrm{F}](t) &\rightarrow [\mathrm{F}]_0^{ss} = P_{total}K_{uf}/(K_{uf}+1),
\end{aligned}
$$
 which is evidently globally asymptotically stable.
\paragraph{In vitro three-state model}
The chemical reactions read
\begin{equation*}\label{A1 in vitro three-state}
	\ce{\mathrm{U} <=>[k^{+}_{ui}][k^{-}_{ui}] I <=>[k^{+}_{if}][k^{-}_{if}] F},
\end{equation*}
whose corresponding kinetic equations can be written as
$$
\begin{aligned}
	\dot{[\mathrm{U}]}(t) & = -k^{+}_{ui}[\mathrm{U}]+k^{-}_{ui}[\mathrm{I}],\label{A1 in vitro three-state dyn1}\\
	\dot{[\mathrm{F}]}(t) & = k^{+}_{if}[\mathrm{I}]-k^{-}_{if}[\mathrm{F}],\label{A1 in vitro three-state dyn2}\\
	[\mathrm{I}](t)& = P_{total} - [\mathrm{U}](t) - [\mathrm{F}](t)\label{A1 in vitro three-state dyn3}.
\end{aligned}
$$
The last equation assumes the mass conservation law of proteins. Accordingly, the analytical expressions are
\begin{align}
	[\mathrm{U}](t)&=C_1\exp(-\lambda_1t)+C_3\exp(-\lambda_2t)+C_5,\\
	[\mathrm{F}](t)&=C_2\exp(-\lambda_1t)+C_4\exp(-\lambda_2t)+C_6,\\
	[\mathrm{I}](t)& = P_{total} - [\mathrm{U}](t) - [\mathrm{F}](t),
\end{align}
where
$$
\begin{aligned}
	\lambda_{1,2}&=\frac{1}{2}(k^{+}_{ui}+k^{-}_{ui}+k^{+}_{if}+k^{-}_{if} \pm ((k^{+}_{ui}+k^{-}_{ui}+k^{+}_{if}+k^{-}_{if})^{2}\\
	&-4(k^{+}_{ui}k^{+}_{if}+k^{+}_{ui}k^{-}_{if}+k^{-}_{ui}k^{-}_{if}))^{1/2}),\\
	C_{1}&=([\mathrm{F}](0)-\xi_{2}[\mathrm{U}](0)-C_{6}+\xi_{2}C_{5})/(\xi_{1}-\xi_{2}), \\
	C_2&=\xi_1C_1, \\
	C_{3}&=\left([\mathrm{F}](0)-\xi_{1}[\mathrm{U}](0)-C_{6}+\xi_{1}C_{5}\right)/\left(\xi_{2}-\xi_{1}\right), \\
	C_4&=\xi_2C_3, \\
	C_{5}&=P_{total}k^{-}_{ui}k^{-}_{if}/(k^{+}_{ui}k^{+}_{if}+k^{+}_{ui}k^{-}_{if}+k^{-}_{ui}k^{-}_{if}), \\
	C_6&=P_{total}k^{+}_{ui}k^{+}_{if}/(k^{+}_{ui}k^{+}_{if}+k^{+}_{ui}k^{-}_{if}+k^{-}_{ui}k^{-}_{if}),\\
	\xi_1&=(\lambda_1-k^{+}_{ui}-k^{-}_{ui})/k^{-}_{ui},\\
	\xi_2&=(\lambda_2-k^{+}_{ui}-k^{-}_{ui})/k^{-}_{ui}.
\end{aligned}
$$
Due to $Re(\lambda_{1,2}) > 0$, we have
$$
\begin{aligned}
	[\mathrm{U}](t) &\rightarrow [\mathrm{U}]^{ss}_0= P_{total}/(1+K_{ui}+K_{ui}K_{if}),\\
	[\mathrm{I}](t) &\rightarrow [\mathrm{I}]_0^{ss} = P_{total}K_{ui}/(1+K_{ui}+K_{ui}K_{if}),\\
	[\mathrm{F}](t) &\rightarrow [\mathrm{F}]_0^{ss} = P_{total}K_{ui}K_{if}/(1+K_{ui}+K_{ui}K_{if}),
\end{aligned}
$$
which is evidently globally asymptotically stable. It is worth noting that when $K_{if}=0$, it degenerates into the steady state of the two-state model.

\paragraph{In vivo two-state model}
The chemical reactions are
\begin{equation*}\label{A1 in vivo two-state}
	\ce{$\emptyset$ <=>[k^{+}_{g}][k^{-}_{g}] U  <=>[k^{+}_{uf}][k^{-}_{uf}] F}.
\end{equation*}
Correspondingly, the kinetic model can be written as
$$\begin{aligned}
	\dot{[\mathrm{U}]}(t) & = -k^{+}_{uf}[\mathrm{U}]+k^{-}_{uf}[\mathrm{F}]-k^{-}_{g}[\mathrm{U}]+k^{+}_{g},\label{A1 in vivo two-state dyn1}\\
	\dot{[\mathrm{F}]}(t) & = k^{+}_{uf}[\mathrm{U}]-k^{-}_{uf}[\mathrm{F}]\label{A1 in vivo two-state dyn2}.
\end{aligned}$$
As this is an open system, there is no mass conservation, thus $[\mathrm{U}](t)+[\mathrm{F}](t)=P_{total}$ does not apply.

Again, we can obtain the analytical expressions
\begin{align}
	[\mathrm{U}](t) & =C_{11}\exp(\lambda_{1} t)+C_{12}\exp(\lambda_{2} t) + K_{g}, \label{vivo-two-state d1}\\
	[\mathrm{F}](t) & =C_{21}\exp(\lambda_{1} t)+C_{22}\exp(\lambda_{2} t) -K_{g}K_{uf} \label{vivo-two-state d2},
\end{align}
where
\begin{align*}
	\lambda_{1,2}& =\frac{1}{2}\left(-(k^{+}_{uf}+k^{-}_{uf}+k^{-}_{g})\pm \sqrt{(k^{+}_{uf}+k^{-}_{uf}+k^{-}_{g})^2-4k^{-}_{uf}k^{-}_{g}} \right),\\
	D_1&=[\mathrm{U}](0)-K_{g}, \\
	D_2&=[\mathrm{F}](0)-K_{g}K_{uf},\\
	C_{11}& =(k^{-}_{uf}D_2-(k^{+}_{uf}+k^{-}_{g}+\lambda_{2})D_1)/(\lambda_{1}-\lambda_{2}), \\
	C_{12}& =(-k^{-}_{uf}D_2+(k^{+}_{uf}+k^{-}_{g}+\lambda_{1})D_1)/(\lambda_{1}-\lambda_{2}), \\
	C_{21}& =(k^{+}_{uf}D_1+(k^{+}_{uf}+k^{-}_{g}+\lambda_{1})D_2)/(\lambda_{1}-\lambda_{2}), \\
	C_{22}& = (-k^{+}_{uf}D_1-(k^{+}_{uf}+k^{-}_{g}+\lambda_{2})D_2)/(\lambda_{1}-\lambda_{2}).
\end{align*}
Since $Re(\lambda_{1,2}) < 0$, we have
$$
\begin{aligned}
	[\mathrm{U}](t) &\rightarrow [\mathrm{U}]^{ss}_0= K_{g},\\
	[\mathrm{F}](t) &\rightarrow [\mathrm{F}]^{ss}_0 = K_{g}K_{uf},
\end{aligned}
$$
which is globally asymptotically stable.

\paragraph{In vivo three-state model}
The chemical reactions read
\begin{equation*}\label{A1 in vivo three-state}
	\ce{$\emptyset$ <=>[k^{+}_{g}][k^{-}_{g}] U <=>[k^{+}_{ui}][k^{-}_{ui}] I <=>[k^{+}_{if}][k^{-}_{if}] F},
\end{equation*}
whose corresponding kinetic equations are
$$
\begin{aligned}
	\dot{[\mathrm{U}]}(t) & = -k^{+}_{ui}[\mathrm{U}]+k^{-}_{ui}[\mathrm{I}]- k^{-}_{g}[\mathrm{U}] + k^{+}_{g},\label{A1 in vivo three-state dyn1}\\
	\dot{[\mathrm{I}]}(t)& =k^{+}_{ui}[\mathrm{U}]-(k^{-}_{ui}+k^{+}_{if})[\mathrm{I}]+k^{-}_{if}[\mathrm{F}],\label{A1 in vivo three-state dyn2}\\
	\dot{[\mathrm{F}]}(t) & = k^{+}_{if}[\mathrm{I}]-k^{-}_{if}[\mathrm{F}]\label{A1 in vivo three-state dyn3}.
\end{aligned}
$$
Again, there is no conservation law for proteins.

After calculation, we obtain the following analytical expressions
\begin{align}
	&\begin{aligned}
	[\mathrm{U}](t)=&C_{11}\exp(-\lambda_1t)+C_{12}\exp(-\lambda_2t)\\&+C_{13}\exp(-\lambda_3t)+K_{g},
	\end{aligned}\\
	&\begin{aligned}
	[\mathrm{I}](t)=&C_{21}\exp(-\lambda_1t)+C_{22}\exp(-\lambda_2t)\\
	&+C_{23}\exp(-\lambda_3t)+K_{g}K_{ui},\end{aligned}\\
	&\begin{aligned}
	[\mathrm{F}](t)=&C_{31}\exp(-\lambda_1t)+C_{32}\exp(-\lambda_2t)\\&+C_{33}\exp(-\lambda_3t)+K_{g}K_{ui}K_{if},
	\end{aligned}
\end{align}
where $\lambda_{1,2,3}$ are the eigenvalues of the matrix
$$A=\left(
\begin{array}{lll}
	-(k^{+}_{ui}+k^{-}_{g})&k^{-}_{ui}&0\\
	k^{+}_{ui}&-(k^{-}_{ui}+k^{+}_{if})&k^{-}_{if}\\
	0&k^{+}_{if}&-k^{-}_{if}
\end{array}\right).
$$
$C_{i}=(C_{1i},C_{21i},C_{3i})^{T}, i=1,2,3$, are the eigenvectors of the matrix A belonging to the eigenvalue $\lambda_{i}$. 

Subsequently, by verifying $Re(\lambda_{1,2,3}) < 0$, we are able to show that
$$
\begin{aligned}
	[\mathrm{U}](t) &\rightarrow [\mathrm{U}]_0^{ss} = K_{g},\\
	[\mathrm{I}](t) &\rightarrow [\mathrm{I}]_0^{ss} = K_{g}K_{ui},\\
	[\mathrm{F}](t) &\rightarrow [\mathrm{F}]_0^{ss}= K_{g}K_{ui}K_{if},
\end{aligned}
$$
and the steady state is globally asymptotically stable. It is worth noting that when $K_{if}=0$, it degenerates into the steady state of the two-state model.

\subsection{Steady states in the presence of stabilizers}
\label{Appendix A2}
We will discuss the steady states of protein folding process in the presence of stabilizers. When stabilizers are present, for both in vitro and in vivo conditions, the two-state model includes additional reversible chemical reactions
$$\ce{F + S <=>[k^{+}_{fs}][k^{-}_{fs}] FS},$$
while for the three-state model, chemical reactions $$\ce{I + S <=>[k^{+}_{is}][k^{-}_{is}] IS}$$ need to be taken into consideration.

\paragraph{In vitro two-state model}
According to the corresponding chemical reactions, we can write down the kinetic equations as
$$
\begin{aligned}
	\dot{[\mathrm{U}]}(t) & = -k^{+}_{uf}[\mathrm{U}]+k^{-}_{uf}[\mathrm{F}],\label{in vitro two-state stab dyn1}\\
	\dot{[\mathrm{F}]}(t) & = k^{+}_{uf}[\mathrm{U}]-k^{-}_{uf}[\mathrm{F}]-k^{+}_{fs}[\mathrm{F}][\mathrm{S}]+k^{-}_{fs}[\mathrm{FS}],\label{in vitro two-state stab dyn2}\\
	\dot{[\mathrm{FS}]}(t) & = k^{+}_{fs}[\mathrm{F}][\mathrm{S}]-k^{-}_{fs}[\mathrm{FS}],\label{in vitro two-state stab dyn3}\\
	\dot{[\mathrm{S}]}(t) & = -k^{+}_{fs}[\mathrm{F}][\mathrm{S}]+k^{-}_{fs}[\mathrm{FS}].\label{in vitro two-state stab dyn4}
\end{aligned}
$$

Clearly, this constitutes a conservative system, satisfying $[\mathrm{U}](t)+[\mathrm{F}](t)+[\mathrm{FS}](t)=P_{total}$ and $[\mathrm{FS}](t)+[\mathrm{S}](t) = S_{total}$. Thus, we can rearrange the equations to yield,
\begin{align}
	&\dot{[\mathrm{U}]}(t)  = -k^{+}_{uf}[\mathrm{U}]+k^{-}_{uf}[\mathrm{F}],\label{in vitro two-state stab dyn5}\\
	&\begin{aligned}
	\dot{[\mathrm{F}]}(t) =& k^{+}_{uf}[\mathrm{U}]-k^{-}_{uf}[\mathrm{F}]+k^{-}_{fs}(P_{total}-[\mathrm{U}]-[\mathrm{F}])\\&-k^{+}_{fs}[\mathrm{F}]([\mathrm{U}]+[\mathrm{F}]-P_{total}+S_{total}),
	\end{aligned}\label{in vitro two-state stab dyn6}
\end{align}
which enables us to obtain the steady state,
$$
\begin{aligned}
	[\mathrm{U}](t) &\rightarrow [\mathrm{U}]^{ss}  = \frac{1}{2}(\sqrt{\eta_{1}^2+4\kappa_{1} P_{total}}-\eta_{1})/(1+K_{uf}),\\
	[\mathrm{F}](t) &\rightarrow [\mathrm{F}]^{ss}  = K_{uf}[\mathrm{U}]^{ss},\\
	[\mathrm{FS}](t) &\rightarrow [\mathrm{FS}]^{ss} = P_{total} - [\mathrm{U}]^{ss} - [\mathrm{F}]^{ss},\\
	[\mathrm{S}](t) &\rightarrow [\mathrm{S}]^{ss}  = S_{total}-P_{total}+[\mathrm{U}]^{ss}+ [\mathrm{F}]^{ss},
\end{aligned}
$$
where $\kappa_{1} = (1+K_{uf})/(K_{uf}K_{fs})$ and $\eta_{1} = S_{total}-P_{total}+\kappa_{1}$.

It is noted that when $S_{total}=0$, $[\mathrm{U}]^{ss}$ degenerates into $[\mathrm{U}]^{ss}_0$. The total amount of folded proteins is denoted by $[\mathrm{FSF}]$, resulting in $[\mathrm{FSF}] = [\mathrm{F}] + [\mathrm{FS}]$ and $[\mathrm{FSF}](t) \rightarrow [\mathrm{FSF}]^{ss} = P_{total} - [\mathrm{U}]^{ss}$.

To assess the stability of the steady state, we perform a variable substitution $x=[\mathrm{U}]-[\mathrm{U}]^{ss}, y= [\mathrm{F}]-[\mathrm{F}]^{ss}$. The resulting linearized equations read
$$\begin{aligned}
	&\dot{x}(t) = -k^{+}_{uf}x +k^{-}_{uf}y,\\
	&\begin{aligned}
	\dot{y}(t) =& (k^{+}_{uf}-k^{+}_{fs}[\mathrm{F}]^{ss}-k^{-}_{fs})x\\&+(-k^{-}_{uf}-k^{+}_{fs}[\mathrm{F}]^{ss}-k^{-}_{fs}-k^{+}_{fs}[\mathrm{U}]^{ss})y
	\end{aligned}
\end{aligned}$$
Because the eigenvalues $\lambda_{1,2}$ corresponding to the matrix,
$$A=\left(
\begin{array}{ll}
	-k^{+}_{uf}&k^{-}_{uf}\\
	k^{+}_{uf}-k^{+}_{fs}[\mathrm{F}]^{ss}-k^{-}_{fs}&-k^{-}_{uf}-k^{+}_{fs}[\mathrm{F}]^{ss}-k^{-}_{fs}-k^{+}_{fs}[\mathrm{U}]^{ss}
\end{array}\right),
$$
satisfy $\lambda_{1}+\lambda_{2}<0$ and $\lambda_{1}\lambda_{2}>0$, it follows that $\lambda_{1}<0$ and $\lambda_{2}<0$. Therefore,  the static state $([\mathrm{U}]^{ss},[\mathrm{F}]^{ss})$ is locally asymptotically stable.

\paragraph{In vitro three-state model}
In this case, the kinetic equations read
$$
\begin{aligned}
	&\dot{[\mathrm{U}]}(t)  = -k^{+}_{ui}[\mathrm{U}]+k^{-}_{ui}[\mathrm{I}],\label{in vitro three-state stab dyn1}\\
	&\dot{[\mathrm{F}]}(t)  = k^{+}_{if}[\mathrm{I}]-k^{-}_{if}[\mathrm{F}],\label{in vitro three-state stab dyn2}\\
	&\begin{aligned}
	\dot{[\mathrm{I}]}(t)  =& k^{+}_{ui}[\mathrm{U}]+k^{-}_{if}[\mathrm{F}]\\&-(k^{-}_{ui}+k^{+}_{if})[\mathrm{I}]-k^{+}_{is}[\mathrm{I}][\mathrm{S}]+k^{-}_{is}[\mathrm{IS}],\label{in vitro three-state stab dyn3}
	\end{aligned}\\
	&\dot{[\mathrm{IS}]}(t)  = k^{+}_{is}[\mathrm{I}][\mathrm{S}]-k^{-}_{is}[\mathrm{IS}],\label{in vitro three-state stab dyn4}\\
	&\dot{[\mathrm{S}]}(t)  = -k^{+}_{is}[\mathrm{I}][\mathrm{S}]+k^{-}_{is}[\mathrm{IS}].\label{in vitro three-state stab dyn5}
\end{aligned}
$$
Clearly, it constitutes a conservative system, satisfying $[\mathrm{U}](t)+[\mathrm{F}](t)+[\mathrm{I}](t)+[\mathrm{IS}](t)=P_{total}$ and $[\mathrm{IS}](t)+[\mathrm{S}](t) = S_{total}$. Correspondingly, we can rearrange equations to yield,
\begin{align}
	&\dot{[\mathrm{U}]}(t) = -k^{+}_{ui}[\mathrm{U}]+k^{-}_{ui}[\mathrm{I}],\label{in vitro three-state stab dyn6}\\
	&\dot{[\mathrm{F}]}(t) = k^{+}_{if}[\mathrm{I}]-k^{-}_{if}[\mathrm{F}],\label{in vitro three-state stab dyn7}\\
	&\begin{aligned}
	\dot{[\mathrm{I}]}(t) =& k^{+}_{ui}[\mathrm{U}]+k^{-}_{if}[\mathrm{F}]-(k^{-}_{ui}+k^{+}_{if})[\mathrm{I}]\\
	&-k^{+}_{is}[\mathrm{I}](S_{total}-P_{total}+[\mathrm{U}]+[\mathrm{F}]+[\mathrm{I}])\\
	&+k^{-}_{is}(P_{total}-[\mathrm{U}]-[\mathrm{F}]-[\mathrm{I}]),\end{aligned}\label{in vitro three-state stab dyn8}
\end{align}
enabling us to solve for the steady state,
$$
\begin{aligned}
	[\mathrm{U}](t) &\rightarrow [\mathrm{U}]^{ss} = \frac{1}{2}(\sqrt{\eta_{2}^2+4\kappa_{2} P_{total}}-\eta_{2})/(1+K_{ui}+K_{ui}K_{if}),\\
	[\mathrm{I}](t) &\rightarrow [\mathrm{I}]^{ss}= K_{ui}[\mathrm{U}]^{ss},\\
	[\mathrm{F}](t) &\rightarrow [\mathrm{F}]^{ss}= K_{ui}K_{if}[\mathrm{U}]^{ss},\\
	[\mathrm{IS}](t) &\rightarrow [\mathrm{IS}]^{ss}  = P_{total} - [\mathrm{U}]^{ss} - [\mathrm{I}]^{ss} - [\mathrm{F}]^{ss},\\
	[\mathrm{S}](t) &\rightarrow [\mathrm{S}]^{ss}  = S_{total}-P_{total}+[\mathrm{U}]^{ss}+[\mathrm{I}]^{ss}+ [\mathrm{F}]^{ss},
\end{aligned}
$$
where $\kappa_{2}= (1+K_{ui}+K_{ui}K_{if})/(K_{ui}K_{fs})$, and $\eta_{2} = S_{total}-P_{total}+\kappa_{2}$.

It is noted that when $S_{total}=0$, $[\mathrm{U}]^{ss}$ degenerates into $[\mathrm{U}]^{ss}_0$; when $K_{if}=0$, it degenerates into the steady state of the two-state model.


\paragraph{In vivo two-state model}
We write down the kinetic equations,
$$
\begin{aligned}
	\dot{[\mathrm{U}]}(t) & = -k^{+}_{uf}[\mathrm{U}]+k^{-}_{uf}[\mathrm{F}]- k^{-}_{g}[\mathrm{U}] +k^{+}_{g},\label{in vivo two-state stab dyn1}\\
	\dot{[\mathrm{F}]}(t) & = k^{+}_{uf}[\mathrm{U}]-k^{-}_{uf}[\mathrm{F}]-k^{+}_{fs}[\mathrm{F}][\mathrm{S}]+k^{-}_{fs}[\mathrm{FS}],\label{in vivo two-state stab dyn2}\\
	\dot{[\mathrm{FS}]}(t) & = k^{+}_{fs}[\mathrm{F}][\mathrm{S}]-k^{-}_{fs}[\mathrm{FS}],\label{in vivo two-state stab dyn3}\\
	\dot{[\mathrm{S}]}(t) & = -k^{+}_{fs}[\mathrm{F}][\mathrm{S}]+k^{-}_{fs}[\mathrm{FS}],\label{in vivo two-state stab dyn4}
\end{aligned}
$$
For stabilizers, being a conservative system, we have $[\mathrm{FS}](t)+[\mathrm{S}](t) = S_{total}$. So that, we can rearrange equations to yield,
\begin{align}
	&\dot{[\mathrm{U}]}(t)  = -k^{+}_{uf}[\mathrm{U}]+k^{-}_{uf}[\mathrm{F}]- k^{-}_{g}[\mathrm{U}] +k^{+}_{g},\label{in vivo two-state stab dyn5}\\
	&\begin{aligned}
	\dot{[\mathrm{F}]}(t) & = k^{+}_{uf}[\mathrm{U}]-k^{-}_{uf}[\mathrm{F}]-k^{+}_{fs}[\mathrm{F}][\mathrm{S}]\\&+k^{-}_{fs}( S_{total}-[\mathrm{S}] ),\end{aligned}\label{in vivo two-state stab dyn6}\\
	&\dot{[\mathrm{S}]}(t) = -k^{+}_{fs}[\mathrm{F}][\mathrm{S}]+k^{-}_{fs}( S_{total}-[\mathrm{S}] ),\label{in vivo two-state stab dyn7}
\end{align}
enabling us to solve for the steady state,
\begin{align*}
	[\mathrm{U}](t) &\rightarrow [\mathrm{U}]^{ss} = K_{g},\\
	[\mathrm{F}](t) &\rightarrow [\mathrm{F}]^{ss} = K_{uf}K_{g},\\
	[\mathrm{S}](t) &\rightarrow [\mathrm{S}]^{ss} = S_{total}/(1+K_{fs}K_{uf}K_{g}),\\
	[\mathrm{FS}](t) &\rightarrow [\mathrm{FS}]^{ss}= S_{total}K_{fs}K_{uf}K_{g}/(1+K_{fs}K_{uf}K_{g}).
\end{align*}
The total amount of folded proteins gives $$[\mathrm{FSF}](t) \rightarrow [\mathrm{FSF}]^{ss} = K_{uf}K_{g} + S_{total}K_{fs}K_{uf}K_{g}/(1+K_{fs}K_{uf}K_{g}).$$

By performing a variable substitution $x=[\mathrm{U}](t)-[\mathrm{U}]^{ss}, y= [\mathrm{F}](t)-[\mathrm{F}]^{ss}$ and $z= [\mathrm{S}](t)-[\mathrm{S}]^{ss}$, we can examine the stability of the resulting linearized equations
\begin{align*}
	\dot{x}(t) & = -(k^{+}_{uf}+k^{-}_{g})x +k^{-}_{uf}y,\\
	\dot{y}(t) & = k^{+}_{uf}x+(-k^{-}_{uf}-k^{+}_{fs}[\mathrm{S}]^{ss})y+(-k^{-}_{fs}-k^{+}_{fs}[\mathrm{F}]^{ss})z,\\
	\dot{z}(t) & = -k^{+}_{fs}[\mathrm{S}]^{ss}y+(-k^{-}_{fs}-k^{+}_{fs}[\mathrm{F}]^{ss})z.
\end{align*}
Let
$$f_{A}(\lambda) = \det(\lambda I - A) = a_{0}\lambda^3 + a_{1}\lambda^2 + a_{2}\lambda + a_{3}$$
be the characteristic polynomial of the matrix $$A=\left(
\begin{array}{lll}
	-(k^{+}_{uf}+k^{-}_{g})& k^{-}_{uf}& 0\\
	k^{+}_{uf}&-(k^{-}_{uf}+k^{+}_{fs}[\mathrm{S}]^{ss})&-(k^{-}_{fs}+k^{+}_{fs}[\mathrm{F}]^{ss})\\
	0&-k^{+}_{fs}[\mathrm{S}]^{ss}&-(k^{-}_{fs}+k^{+}_{fs}[\mathrm{F}]^{ss})
\end{array}\right),
$$ where \begin{align*}
	&a_{0}  = 1,\\
	&a_{1}  = k^{+}_{uf}+k^{-}_{g}+k^{-}_{uf}+k^{+}_{fs}[\mathrm{S}]^{ss}+ k^{-}_{fs}+k^{+}_{fs}[\mathrm{F}]^{ss},\\
	&\begin{aligned}
	a_{2}  =& (k^{+}_{uf}+k^{-}_{g})(k^{-}_{uf}+k^{-}_{fs}+k^{+}_{fs}[\mathrm{S}]^{ss}+k^{+}_{fs}[\mathrm{F}]^{ss})\\&+k^{-}_{uf}(k^{+}_{fs}[\mathrm{F}]^{ss}+k^{-}_{fs}-k^{+}_{uf}),\end{aligned}\\
	&a_{3}  = k^{-}_{uf}k^{-}_{g}(k^{+}_{fs}[\mathrm{F}]^{ss}+k^{-}_{fs}).
\end{align*}
From $a_{0}>0,\Delta_{1}=a_{1}>0,a_{3}>0$ and
$$\Delta_{2} =
\Bigg| \begin{array}{ll}
	a_{1}&a_{0}\\
	a_{3}&a_{2}
\end{array}
\Bigg|>0,
$$
we conclude that the real parts of the eigenvalues of matrix $ A $ are all less than zero, indicating a local asymptotic stability.

\paragraph{In vivo three-state model}
The kinetic equations read
$$
\begin{aligned}
	&\dot{[\mathrm{U}]}(t)  = -k^{+}_{ui}[\mathrm{U}]+k^{-}_{ui}[\mathrm{I}]- k^{-}_{g}[\mathrm{U}] +k^{+}_{g},\label{in vivo three-state stab dyn1}\\
	&\begin{aligned}
	\dot{[\mathrm{I}]}(t)& =k^{+}_{ui}[\mathrm{U}]-k^{-}_{ui}[\mathrm{I}] - k^{+}_{if}[\mathrm{I}]+k^{-}_{if}[\mathrm{F}]\\&-k^{+}_{is}[\mathrm{I}][\mathrm{S}]+k^{-}_{is}[\mathrm{IS}], \end{aligned}\label{in vivo three-state stab dyn2}\\
	&\dot{[\mathrm{F}]}(t)  = k^{+}_{if}[\mathrm{I}]-k^{-}_{if}[\mathrm{F}],\label{in vivo three-state stab dyn3}\\
	&\dot{[\mathrm{IS}]}(t)  = k^{+}_{is}[\mathrm{I}][\mathrm{S}]-k^{-}_{is}[\mathrm{IS}],\label{in vivo three-state stab dyn4}\\
	&\dot{[\mathrm{S}]}(t)  = -k^{+}_{is}[\mathrm{I}][\mathrm{S}]+k^{-}_{is}[\mathrm{IS}],\label{in vivo three-state stab dyn5}
\end{aligned}
$$

For stabilizers, being a conservative system, we have $[\mathrm{IS}](t)+[\mathrm{S}](t) = S_{total}$. Thus, we can rearrange equations as
\begin{align}
	&\dot{[\mathrm{U}]}(t)  = -k^{+}_{ui}[\mathrm{U}]+k^{-}_{ui}[\mathrm{I}]- k^{-}_{g}[\mathrm{U}] +k^{+}_{g},\label{in vivo three-state stab dyn6}\\
	&\begin{aligned}
	\dot{[\mathrm{I}]}(t)& =k^{+}_{ui}[\mathrm{U}]-k^{-}_{ui}[\mathrm{I}] - k^{+}_{if}[\mathrm{I}]+k^{-}_{if}[\mathrm{F}]\\& -k^{+}_{is}[\mathrm{I}][\mathrm{S}]+k^{-}_{is}(S_{total}-[\mathrm{S}]),
	\end{aligned}\label{in vivo three-state stab dyn7}\\
	&\dot{[\mathrm{F}]}(t)  = k^{+}_{if}[\mathrm{I}]-k^{-}_{if}[\mathrm{F}],\label{in vivo three-state stab dyn8}\\
	&\dot{[\mathrm{S}]}(t)  = -k^{+}_{is}[\mathrm{I}][\mathrm{S}]+k^{-}_{is}(S_{total}-[\mathrm{S}]),\label{in vivo three-state stab dyn9}
\end{align}
whose steady state is given by
\begin{align*}
	[\mathrm{U}](t) &\rightarrow [\mathrm{U}]^{ss} = K_{g},\\
	[\mathrm{I}](t) &\rightarrow [\mathrm{I}]^{ss} = K_{ui}K_{g},\\
	[\mathrm{F}](t) &\rightarrow [\mathrm{F}]^{ss} = K_{if}K_{ui}K_{g},\\
	[\mathrm{S}](t) &\rightarrow [\mathrm{S}]^{ss} = S_{total}/(1+K_{is}K_{ui}K_{g}),\\
	[\mathrm{IS}](t) &\rightarrow [\mathrm{IS}]^{ss} = S_{total}K_{is}K_{ui}K_{g}/(1+K_{is}K_{ui}K_{g}).
\end{align*}
In addition, the total amount of intermediates is $$[\mathrm{ISI}](t) \rightarrow [\mathrm{ISI}]^{ss} =K_{ui}K_{g} + S_{total}K_{is}K_{ui}K_{g}/(1+K_{is}K_{ui}K_{g}).$$

By performing a variable substitution $x=[\mathrm{U}](t)-[\mathrm{U}]^{ss}, y= [\mathrm{I}](t)-[\mathrm{I}]^{ss}, z = [\mathrm{S}](t)-[\mathrm{S}]^{ss}$ and $u = [\mathrm{F}](t)-[\mathrm{F}]^{ss}$, we examine the stability of the resulting linearized equations,
\begin{align*}
	\dot{x}(t) & = -(k^{+}_{ui}+k^{-}_{g})x +k^{-}_{ui}y,\\
	\dot{y}(t) & = k^{+}_{ui}x-(k^{-}_{ui}+k^{+}_{if}+k^{+}_{is}[\mathrm{S}]^{ss})y-(k^{-}_{is}+k^{+}_{is}[\mathrm{I}]^{ss})z+k^{-}_{if}u,\\
	\dot{z}(t) & = -k^{+}_{is}[\mathrm{S}]^{ss}y-(k^{-}_{is}+k^{+}_{is}[\mathrm{I}]^{ss})z,\\
	\dot{u}(t) & = k^{+}_{if}y-k^{-}_{if}u.
\end{align*}
Let
$$f_{A}(\lambda) = \det(\lambda I - A) = a_{0}\lambda^4 + a_{1}\lambda^3 + a_{2}\lambda^{2} + a_{3}\lambda + a_{4} $$
be the characteristic polynomial of the matrix,  $$A=\left(
\begin{array}{llll}
	-(k^{+}_{ui}+k^{-}_{g})& k^{-}_{ui}& 0&0\\
	k^{+}_{ui}&-(k^{-}_{ui}+k^{+}_{if}+k^{+}_{is}[\mathrm{S}]^{ss})&-(k^{-}_{is}+k^{+}_{is}[\mathrm{I}]^{ss})&k^{-}_{if}\\
	0&-k^{+}_{is}[\mathrm{S}]^{ss}&-(k^{-}_{is}+k^{+}_{is}[\mathrm{I}]^{ss})&0\\
	0&k^{+}_{if}&0&-k^{-}_{if}
\end{array}\right)
$$ where \begin{align*}
	&a_{0}  = 1,\\
	&a_{1}  = k^{+}_{ui}+k^{-}_{ui}+k^{+}_{if}+k^{-}_{if}+k^{-}_{g}+k^{-}_{is}+k^{+}_{is}[\mathrm{S}]^{ss} +k^{+}_{is}[\mathrm{I}]^{ss},\\
	&\begin{aligned}a_{2}  =&k^{+}_{ui}(k^{+}_{if}+k^{+}_{is}[\mathrm{S}]^{ss})+k^{-}_{if}(k^{-}_{ui}+k^{+}_{ui}+k^{+}_{is}[\mathrm{S}]^{ss})\\&+k^{-}_{g}(k^{-}_{ui}+k^{-}_{if}+k^{+}_{if}+k^{+}_{is}[\mathrm{S}]^{ss})\\
	&+(k^{-}_{is}+k^{+}_{is}[\mathrm{I}]^{ss})(k^{+}_{ui}+k^{-}_{ui}+k^{+}_{if}+k^{-}_{if}+k^{-}_{g}),\end{aligned}\\
	&\begin{aligned}a_{3} &=(k^{-}_{is}+k^{+}_{is}[\mathrm{I}]^{ss})\left[k^{-}_{if}(k^{+}_{ui}+k^{-}_{ui}+k^{+}_{if})+k^{-}_{g}(k^{-}_{ui}+k^{-}_{if}+k^{+}_{if})\right]\\&+k^{+}_{is}k^{-}_{if}[\mathrm{S}]^{ss}(k^{+}_{ui}+k^{-}_{g})+k^{-}_{g}k^{-}_{ui}k^{-}_{if},\end{aligned} \\
	&a_{4}  = k^{-}_{ui}k^{-}_{if}k^{-}_{g}(k^{+}_{is}[\mathrm{I}]^{ss}+k^{-}_{is}).
\end{align*}
From $a_{0}>0,\Delta_{1}=a_{1}>0,a_{4}>0$,
$$\Delta_{2} =
\Bigg| \begin{array}{ll}
	a_{1}&a_{0}\\
	a_{3}&a_{2}
\end{array}
\Bigg|>0,
$$
and
$$\Delta_{3} =
\Bigg| \begin{array}{lll}
	a_{1}&a_{0}&0\\
	a_{3}&a_{2}&a_{1}\\
	a_{5}&a_{4}&a_{3}
\end{array}
\Bigg|=\Bigg| \begin{array}{lll}
	a_{1}&a_{0}&0\\
	a_{3}&a_{2}&a_{1}\\
	0&a_{4}&a_{3}
\end{array}
\Bigg|>0,
$$
we conclude that the real parts of the eigenvalues of matrix $ A $ are all less than zero, indicating a local asymptotic stability.

\subsection{Steady state independent of the way of stabilizer addition}
\label{Appendix A3}
In the preceding discussions, we have assumed that all stabilizers have been added at the very beginning. Now, we illustrate, using the in-vitro two-state model as an example, that the steady state is independent of the way how stabilizers are added; it is solely determined by the total amount of stabilizers.

Assume that the initial amount of stabilizers is $\mathrm{S}_0$, and the rate of stabilizer addition over time is represented by the function $u(t)$  $(u(t) \geq 0, \forall t \in \mathbb{R}^+)$. It satisfies the existence of a finite time $T$, such that $\mathrm{S}_0 + \int_{0}^{T} u(t) dt = \mathrm{S}_0 + \int_{0}^{\infty} u(t) dt = S_{total}$.

The kinetic equations considering stabilizer addition are given by
\begin{align*}
	\dot{[\mathrm{U}]}(t) & = -k^{+}_{uf}[\mathrm{U}]+k^{-}_{uf}[\mathrm{F}],\\
	\dot{[\mathrm{F}]}(t) & = k^{+}_{uf}[\mathrm{U}]-k^{-}_{uf}[\mathrm{F}]-k^{+}_{fs}[\mathrm{F}][\mathrm{S}]+k^{-}_{fs}[\mathrm{FS}],\\
	\dot{[\mathrm{FS}]}(t) & = k^{+}_{fs}[\mathrm{F}][\mathrm{S}]-k^{-}_{fs}[\mathrm{FS}],\\
	\dot{[\mathrm{S}]}(t) & = -k^{+}_{fs}[\mathrm{F}][\mathrm{S}]+k^{-}_{fs}[\mathrm{FS}] + u.
\end{align*}
Due to laws of mass conservation, they can be simplied into
\begin{align*}
	&\dot{[\mathrm{U}]}(t)  = -k^{+}_{uf}[\mathrm{U}]+k^{-}_{uf}[\mathrm{F}],\\
	&\begin{aligned}
		\dot{[\mathrm{F}]}(t)  =& k^{+}_{uf}[\mathrm{U}]-k^{-}_{uf}[\mathrm{F}]+k^{-}_{fs}(P_{total}-[\mathrm{U}]-[\mathrm{F}])\\&-k^{+}_{fs}[\mathrm{F}]([\mathrm{U}]+[\mathrm{F}]-P_{total}+\mathrm{S}_{0}+\int_{0}^{t}u(\tau) d\tau ).\end{aligned}
\end{align*}
As we are only interested in the steady state, we can suppose $t>T$ in the above equation and thus obtain
\begin{align*}
	&\dot{[\mathrm{U}]}(t)  = -k^{+}_{uf}[\mathrm{U}]+k^{-}_{uf}[\mathrm{F}],\\
	&\begin{aligned}
		\dot{[\mathrm{F}]}(t) & = k^{+}_{uf}[\mathrm{U}]-k^{-}_{uf}[\mathrm{F}]+k^{-}_{fs}(P_{total}-[\mathrm{U}]-[\mathrm{F}])\\&-k^{+}_{fs}[\mathrm{F}]([\mathrm{U}]+[\mathrm{F}]-P_{total}+S_{total} ).
	\end{aligned}
\end{align*}
This is the same equation we considered before, thus resulting in the same steady state.
\section{Control of the steady state}
\label{Appendix B}
This subsection will analyze four scenarios to determine the total amount of stabilizers required for reaching the desired increased folds $\gamma_2$ or $\gamma_3$.

\paragraph{In vitro two-state model}
According to the results on the stabilizer-constrained steady state in \ref{Appendix A2}, we can obtain $$[\mathrm{FSF}]^{ss}=P_{total}-\frac{1}{2}(\sqrt{\eta_{1}^2+4\kappa_{1} P_{total}}-\eta_{1})/(1+K_{uf}).$$ Comparing with the results without stabilizers from \ref{Appendix A1}, we can derive $$S_{expected} = (\kappa_{1}K_{uf}\gamma_{2})/(1-K_{uf}\gamma_{2})+P_{total}K_{uf}\gamma_{2}.$$
When $\kappa_{1}/({1-K_{uf}\gamma_{2}}) \ll P_{total}$, or equivalently, $\gamma_{2} \ll (1-\kappa_{1}/P_{total})/K_{uf}$, which occurs when $K_{uf}\ll1$ and $K_{uf}K_{fs}P_{total}\gg1 $, for most $\gamma_{2}>0$, we can approximate the total concentration of  stabilizers as $$S_{expected} \approx P_{total}K_{uf}\gamma_{2}.$$
\paragraph{In vitro three-state model}
In this case, we have $$[\mathrm{ISI}]^{ss}=P_{total}-\frac{1}{2}(\sqrt{\eta_{2}^2+4\kappa_{2} P_{total}}-\eta_{2})/(1+\widetilde{K_{uf}}),$$
where $\widetilde{K_{uf}} = K_{ui}/(1+K_{ui}K_{if}).$ Furthermore, $$S_{expected} = (\kappa_{2}\widetilde{K_{uf}}\gamma_{3})/(1-\widetilde{K_{uf}}\gamma_{3})+P_{total}\widetilde{K_{uf}}\gamma_{3}.$$
When $\kappa_{2}/({1-\widetilde{K_{uf}}\gamma_{3}}) \ll P_{total}$, or equivalently, $\gamma_{3} \ll (1-\kappa_{2}/P_{total})/\widetilde{K_{uf}}$, which occurs when $\widetilde{K_{uf}}\ll1$ and $\widetilde{K_{uf}}K_{fs}P_{total}\gg1 $, for most $\gamma_{3}>0$, we can approximate the total concentration of  stabilizers as $$S_{expected} \approx P_{total}\widetilde{K_{uf}}\gamma_{3}.$$

\paragraph{In vivo two-state model}
After simple calculation, we can obtain 
\begin{align*}
[\mathrm{FSF}]^{ss}&=K_{uf}K_{g}+ S_{total}K_{fs}K_{uf}K_{g}/(1+K_{fs}K_{uf}K_{g}),\\
S_{expected} &=\gamma_{2}(K_{uf}K_{g}+1/K_{fs}).
\end{align*}

\paragraph{In vivo three-state model}
In this case, we have
\begin{align*}
[\mathrm{ISI}]^{ss}&=K_{ui}K_{g}+ S_{total}K_{is}K_{ui}K_{g}/(1+K_{is}K_{ui}K_{g}),\\
S_{expected} &=\gamma_{3}(K_{ui}K_{g}+1/K_{is}).
\end{align*}

\section{Optimal control for stabilizing folded proteins}
\label{Appendix C}
We begin by presenting the general formulation of the optimal control problem along with the Pontryagin maximum principle regarding to its necessary conditions. Subsequently, we delineate the corresponding optimal control problems for four distinct scenarios. Finally, we discuss the optimal switching time for the Bang-Bang controls.

\subsection{Optimal control problem and necessary conditions}
\label{Appendix C1}
This section solely considers optimal control problems with fixed terminal time, comprising three components: the objective functional, dynamics influenced by controls, and control constraints. These can be mathematically expressed as:
$$
\begin{aligned}
	\text{Minimize } & J[u(\cdot)] = \phi(X(T)) + \int_{0}^{T} L(X(t), u(t), t) dt \\
	\text{Subject to } & \dot{X}(t) = f(X(t), u(t), t), \quad t \in [0, T] \\
	& X(0) = X_0, \\
	& u(t) \in \mathrm{U}, \quad t \in [0, T]
\end{aligned}
$$
where $ X \in \mathbb{R}^d $ represents the state variable, $ u \in \mathbb{R}^m $ denotes the control variable, $ T $ signifies the terminal time, $ \phi $ represents the terminal cost, $ L $ denotes the running cost, $ f $ signifies the dynamics influenced by the control, $ X(0) $ represents the initial states, and $ \mathrm{U} \subset \mathbb{R}^d$ represents the control constraint set.

Pontryagin's maximum principle offers a necessary condition for this optimal control problem, providing theoretical support for analyzing the solutions of optimal control problems.

\begin{lemma}
	Let $ u^* \in \mathrm{U} $ be a bounded, measurable, admissible control that minimizes the objective functional described above, and $ X^* $ be the corresponding state trajectory.
	
	Define the Hamiltonian, $$ H(X,u,p,t) = p^T f(X,u,t) - L(X,u,t), $$ where $ p\in \mathbb{R}^d $. Then there exists an absolutely continuous process $ p^*(t) $ such that:
	
	$$
	\begin{aligned}
		&\dot{X}^*(t) = \frac{\partial H(X^*(t),u^*(t),p^*(t),t)}{\partial p}, \quad X^*(0) = x_0 \\
		&\dot{p}^*(t) = -\frac{\partial H(X^*(t),u^*(t),p^*(t),t)}{\partial X}, \quad p^*(T) = -\frac{\partial\phi(X^*(T))}{\partial X} \\
		&H(X^*(t),u^*(t),p^*(t),t) \geq H(X^*(t),u(t),p^*(t),t), \\
		&\forall u \in \mathrm{U} \quad \text{and} \quad \text{a.e. } t \in [0,T],
	\end{aligned}
	$$
\end{lemma}
Therefore, we can seek solutions for the optimal control by constructing such a system.

\subsection{Optimal control for stabilizing proteins at speficied functional states}
\label{Appendix C2}
Because the constraints and objective function for the optimal control of stabilizing proteins at certain specified functioal state are similar across all scenarios considered, we will present them in a unified manner. Subsequently, we will provide the controlled dynamics for each scenario and the corresponding necessary conditions.

\paragraph{Objective functional and constraints}
The objective functional for the two-state model reads
\begin{equation*}
	J[u(\cdot)] = - w_{1}([\mathrm{FSF}](T))+w_{2}\int_{0}^{T}u(\tau)d \tau.
\end{equation*}
Here $[\mathrm{FSF}] = [\mathrm{FS}]+[\mathrm{F}].$ T denotes the finite terminal time considered in the analysis. The function $u(t)$ represents the rate of stabilizer addition over time. Finally, $ w_{1} $ and $ w_{2} $ respectively represent the weighting coefficients for folded proteins and stabilizers. Denoting $ w_{Tol} = w_{1}/w_{2}$, an equivalent objective function is \begin{equation}\label{Appendix objective functional two-state}
	J[u(\cdot)] = - w_{Tol}([\mathrm{FSF}](T))+\int_{0}^{T}u(\tau)d \tau.
\end{equation}
This is the exact form considered in the main text.

Similarly, for the three-state model, we can establish the objective functional,
\begin{equation}\label{Appendix objective functional three-state}
	J[u(\cdot)] = - w_{Tol}([\mathrm{ISI}](T))+\int_{0}^{T}u(\tau)d \tau.
\end{equation}

In practice, we impose a constraint on the additin rate $ u(t) $ such that
\begin{equation}\label{Appendix constraints}
	0 \leq u(t)\leq u_{\max},\quad \forall  t \in [0, T],
\end{equation}
where $ u_{\max} $ represents the maximum rate of stabilizer addition.

\paragraph{In vitro two-state model}
we present the kinetics of protein folding affected by stabilizer addition as
$$
\begin{aligned}
	\dot{[\mathrm{U}]}(t) & = -k^{+}_{uf}[\mathrm{U}]+k^{-}_{uf}[\mathrm{F}],\\
	\dot{[\mathrm{F}]}(t) & = k^{+}_{uf}[\mathrm{U}]-k^{-}_{uf}[\mathrm{F}]-k^{+}_{fs}[\mathrm{F}][\mathrm{S}]+k^{-}_{fs}[\mathrm{FS}],\\
	\dot{[\mathrm{FS}]}(t) & = k^{+}_{fs}[\mathrm{F}][\mathrm{S}]-k^{-}_{fs}[\mathrm{FS}],\\
	\dot{[\mathrm{S}]}(t) & = -k^{+}_{fs}[\mathrm{F}][\mathrm{S}]+k^{-}_{fs}[\mathrm{FS}]+u.
\end{aligned}
$$

By applying laws of mass conservation, which state that $[\mathrm{U}] + [\mathrm{FSF}] = P_{total}$, we can transform the objective functional in  \eqref{Appendix objective functional two-state} into \begin{equation*}
	J[u(\cdot)] =  w_{Tol}([\mathrm{U}](T))+\int_{0}^{T}u(\tau)d \tau.
\end{equation*}
Then by utilizing the Pontryagin's maximum principle, we can obtain the following necessary conditions,
\begin{align}
	&\dot{[\mathrm{U}]}(t) = -k^{+}_{uf}[\mathrm{U}]+k^{-}_{uf}[\mathrm{F}],\nonumber \\
	&\begin{aligned}
	\dot{[\mathrm{F}]}(t) & = (k^{+}_{uf}-k^{-}_{fs})[\mathrm{U}]-(k^{-}_{uf}+k^{-}_{fs})[\mathrm{F}]\\&-k^{+}_{fs}[\mathrm{F}][\mathrm{S}]+k^{-}_{fs}P_{total},\end{aligned}\nonumber\\
	&\dot{[\mathrm{S}]}(t) =-k^{-}_{fs}[\mathrm{U}] -k^{-}_{fs}[\mathrm{F}] -k^{+}_{fs}[\mathrm{F}][\mathrm{S}]+k^{-}_{fs}P_{total} + u, \nonumber\\
	&\dot{p_{1}}(t) = k^{+}_{uf}p_{1}-(k^{+}_{uf}-k^{-}_{fs})p_{2}+k^{-}_{fs}p_{3},\label{in vitro two-state bvp1}\\
	&\begin{aligned}
		\dot{p_{2}}(t)  =&-k^{-}_{uf}p_{1}+(k^{-}_{uf}+k^{-}_{fs}+k^{+}_{fs}[\mathrm{S}])p_{2}\\&+(k^{-}_{fs}+k^{+}_{fs}[\mathrm{S}])p_{3},
	\end{aligned}\label{in vitro two-state bvp2}\\
	&\dot{p_{3}}(t)  = k^{+}_{fs}[\mathrm{F}]p_{2}+k^{+}_{fs}[\mathrm{F}]p_{3},\label{in vitro two-state bvp3}\\
	&(p_{3}-1)u  = \max_{a\in [0, u_{\max}]}((p_{3}-1)a),\label{in vitro two-state bvp4}\\
	&[\mathrm{U}](0) = P_{total}, [\mathrm{F}](0) = [\mathrm{S}](0) = 0, \mathbf{p}(T)=(-w_{Tol},0,0).\nonumber
\end{align}
Here we omit the superscript "*" indicating optimality (for example, $ u^{*} $ denotes the optimal control).

It is worth noting that, based on \eqref{in vitro two-state bvp4}, it is straightforward to see that the optimal control $u(t)$ is a Bang-Bang control, as long as $ p_{3}-1 = 0 $ has finite roots. Thus we refer $p_3$ as the key adjoint state, denoted by $\tilde{p}$.

\paragraph{In vitro three-state model}
In this case, the kinetic model is given by
\begin{equation}\label{Appendix vitro three-state}
\begin{aligned}
	&\dot{[\mathrm{U}]}(t)  = -k^{+}_{ui}[\mathrm{U}]+k^{-}_{ui}[\mathrm{I}],\\
	&\dot{[\mathrm{F}]}(t)  = k^{+}_{if}[\mathrm{I}]-k^{-}_{if}[\mathrm{F}],\\
	&\begin{aligned}
		\dot{[\mathrm{I}]}(t)  =& k^{+}_{ui}[\mathrm{U}]+k^{-}_{if}[\mathrm{F}]\\&-(k^{-}_{ui}+k^{+}_{if})[\mathrm{I}]-k^{+}_{is}[\mathrm{I}][\mathrm{S}]+k^{-}_{is}[\mathrm{IS}],
	\end{aligned}\\
	&\dot{[\mathrm{IS}]}(t)  = k^{+}_{is}[\mathrm{I}][\mathrm{S}]-k^{-}_{is}[\mathrm{IS}],\\
	&\dot{[\mathrm{S}]}(t)  = -k^{+}_{is}[\mathrm{I}][\mathrm{S}]+k^{-}_{is}[\mathrm{IS}]+u.
\end{aligned}
\end{equation}

According to laws of mass conservation and the Pontryagin's maximum principle, we can obtain the following necessary conditions,
$$
\begin{aligned}
	&\dot{[\mathrm{U}]}(t)  = -k^{+}_{ui}[\mathrm{U}]+k^{-}_{ui}[\mathrm{I}],\\
	&\begin{aligned}
		\dot{[\mathrm{I}]}(t) & = (k^{+}_{ui}-k^{-}_{if})[\mathrm{U}]-(k^{-}_{ui}+k^{+}_{if}+k^{-}_{is})[\mathrm{I}]\\
		&+(k^{-}_{is}-k^{-}_{if})[\mathrm{ISI}]-k^{+}_{is}[\mathrm{I}][\mathrm{S}]+k^{-}_{if}P_{total},
	\end{aligned}
	\\
	&\dot{[\mathrm{ISI}]}(t) = (k^{+}_{ui}-k^{-}_{if})[\mathrm{U}]-(k^{-}_{ui}+k^{+}_{if})[\mathrm{I}]-k^{-}_{if}[\mathrm{ISI}]+k^{-}_{if}P_{total},\\
	&\dot{[\mathrm{S}]}(t) = -k^{-}_{is}[\mathrm{I}]+k^{-}_{is}[\mathrm{ISI}]-k^{+}_{is}[\mathrm{I}][\mathrm{S}]+u,\\
	&\dot{p_{1}}(t) = k^{+}_{ui}p_{1}-(k^{+}_{ui}-k^{-}_{if})(p_{2}+p_{3}),\\
	&\dot{p_{2}}(t) =-k^{-}_{ui}p_{1}+(k^{-}_{ui}+k^{+}_{if})(p_{2}+p_{3})+(k^{-}_{is}+k^{+}_{is}[\mathrm{S}])(p_{2}+p_{4}),\\
	&\dot{p_{3}}(t) =k^{-}_{if}(p_{2}+p_{3})-k^{-}_{is}(p_{2}+p_{4}),\\
	&\dot{p_{4}}(t) = k^{+}_{is}[\mathrm{I}](p_{2}+p_{4}),\\
	&(p_{4}-1)u = \max_{a\in [0, u_{\max}]}((p_{4}-1)a),\\
	&[\mathrm{U}](0) =P_{total}, [\mathrm{I}](0) = [\mathrm{ISI}](0) = [\mathrm{S}](0) = 0, \\
	& \mathbf{p}(T)=(0,0,w_{Tol},0).
\end{aligned}
$$

Similarly, the optimal control $u(t)$ is a Bang-Bang control, as long as the key adjoint state $ p_{4}-1 = 0 $ has finite roots.

\paragraph{In vivo two-state model}
Considering the following kinetic equations for the in-vivo two-state model,
$$
\begin{aligned}
	\dot{[\mathrm{U}]}(t) & = -k^{+}_{uf}[\mathrm{U}]+k^{-}_{uf}[\mathrm{F}]- k^{-}_{g}[\mathrm{U}] +k^{+}_{g},\\
	\dot{[\mathrm{F}]}(t) & = k^{+}_{uf}[\mathrm{U}]-k^{-}_{uf}[\mathrm{F}]-k^{+}_{fs}[\mathrm{F}][\mathrm{S}]+k^{-}_{fs}[\mathrm{FS}],\\
	\dot{[\mathrm{FS}]}(t) & = k^{+}_{fs}[\mathrm{F}][\mathrm{S}]-k^{-}_{fs}[\mathrm{FS}],\\
	\dot{[\mathrm{S}]}(t) & = -k^{+}_{fs}[\mathrm{F}][\mathrm{S}]+k^{-}_{fs}[\mathrm{FS}]+u.
\end{aligned}
$$
we can obtain the necessary conditions as
$$
\begin{aligned}
	&\dot{[\mathrm{U}]}(t)  = -k^{+}_{uf}[\mathrm{U}]+k^{-}_{uf}[\mathrm{F}]- k^{-}_{g}[\mathrm{U}] +k^{+}_{g}\\
	&\dot{[\mathrm{F}]}(t)  = k^{+}_{uf}[\mathrm{U}]-k^{-}_{uf}[\mathrm{F}]-k^{+}_{fs}[\mathrm{F}][\mathrm{S}]+k^{-}_{fs}([\mathrm{FSF}]-[\mathrm{F}])\\
	&\dot{[\mathrm{FSF}]}(t)  =  k^{+}_{uf}[\mathrm{U}]-k^{-}_{uf}[\mathrm{F}]\\
	&\dot{[\mathrm{S}]}(t)  = -k^{+}_{fs}[\mathrm{F}][\mathrm{S}]+k^{-}_{fs}([\mathrm{FSF}]-[\mathrm{F}])+u \\
	&\dot{p_{1}}(t)  = (k^{+}_{uf}+k^{-}_{g})p_{1}-k^{+}_{uf}(p_{2}+p_{3})\\
	&\dot{p_{2}}(t)  =-k^{-}_{uf}(p_{1}-p_{2}-p_{3})+(k^{-}_{fs}+k^{+}_{fs}[\mathrm{S}])(p_{2}+p_{4})\\
	&\dot{p_{3}}(t)  = -k^{-}_{fs}(p_{2}+p_{4})\\
	&\dot{p_{4}}(t)  = k^{+}_{fs}[\mathrm{F}](p_{2}+p_{4})\\
	&(p_{4}-1)u  = \max_{a\in [0, u_{\max}]}((p_{4}-1)a)\\
	&[\mathrm{U}](0) =P_{total}, [\mathrm{F}](0)= [\mathrm{FSF}](0) = [\mathrm{S}](0) = 0 \\ &\mathbf{p}(T)=(0,0,w_{Tol},0).\label{OC vivo-two-state stab add-bvp10}
\end{aligned}$$

In this case, $p_4$ acts as the key adjoint state, denoted by $\tilde{p}$.

\paragraph{In vivo three-state model}
Here the kinetic equations for three-state protein folding affected by stabilizer addition are
$$
\begin{aligned}
	&\dot{[\mathrm{U}]}(t)  = -k^{+}_{ui}[\mathrm{U}]+k^{-}_{ui}[\mathrm{I}]- k^{-}_{g}[\mathrm{U}] +k^{+}_{g},\\
	&\begin{aligned}
		\dot{[\mathrm{I}]}(t)& =k^{+}_{ui}[\mathrm{U}]-k^{-}_{ui}[\mathrm{I}] - k^{+}_{if}[\mathrm{I}]+k^{-}_{if}[\mathrm{F}]\\&-k^{+}_{is}[\mathrm{I}][\mathrm{S}]+k^{-}_{is}[\mathrm{IS}], \end{aligned}\\
	&\dot{[\mathrm{F}]}(t)  = k^{+}_{if}[\mathrm{I}]-k^{-}_{if}[\mathrm{F}],\\
	&\dot{[\mathrm{IS}]}(t)  = k^{+}_{is}[\mathrm{I}][\mathrm{S}]-k^{-}_{is}[\mathrm{IS}],\\
	&\dot{[\mathrm{S}]}(t)  = -k^{+}_{is}[\mathrm{I}][\mathrm{S}]+k^{-}_{is}[\mathrm{IS}]+u,
\end{aligned}
$$

Applying the Pontryagin's maximum principle, we  arrive at

$$
\begin{aligned}
	&\dot{[\mathrm{U}]}(t)  = -k^{+}_{ui}[\mathrm{U}]+k^{-}_{ui}[\mathrm{I}]- k^{-}_{g}[\mathrm{U}] +k^{+}_{g}\\
	&\begin{aligned}\dot{[\mathrm{I}]}(t) =&k^{+}_{ui}[\mathrm{U}]-k^{-}_{ui}[\mathrm{I}] - k^{+}_{if}[\mathrm{I}]+k^{-}_{if}[\mathrm{F}]\\& -k^{+}_{is}[\mathrm{I}][\mathrm{S}]+k^{-}_{is}([\mathrm{ISI}]-[\mathrm{I}])\end{aligned}\\
	&\dot{[\mathrm{F}]}(t)  = k^{+}_{if}[\mathrm{I}]-k^{-}_{if}[\mathrm{F}]\\
	&\dot{[\mathrm{ISI}]}(t)  = k^{+}_{ui}[\mathrm{U}]-k^{-}_{ui}[\mathrm{I}] - k^{+}_{if}[\mathrm{I}]+k^{-}_{if}[\mathrm{F}]\\
	&\dot{[\mathrm{S}]}(t)  = -k^{+}_{is}[\mathrm{I}][\mathrm{S}]+k^{-}_{is}([\mathrm{ISI}]-[\mathrm{I}])+u \\
	&\dot{p_{1}}(t)  = k^{+}_{ui}(p_{1}-p_{2}-p_{4})+k^{-}_{g}p_{1}\\
	&\begin{aligned}\dot{p_{2}}(t)  =&-k^{-}_{ui}(p_{1}-p_{2}-p_{4})+k^{+}_{if}(p_{2}-p_{3}+p_{4})\\&+(k^{-}_{is}+k^{+}_{is}[\mathrm{S}])(p_{2}+p_{5})\end{aligned}\\
	&\dot{p_{3}}(t)  = -k^{-}_{if}(p_{2}-p_{3}+p_{4})\\
	&\dot{p_{4}}(t)  = -k^{-}_{is}(p_{2}+p_{5})\\
	&\dot{p_{5}}(t)  = k^{+}_{is}[\mathrm{I}](p_{2}+p_{5})\\
	&(p_{5}-1)u  = \max_{a\in [0, u_{\max}]}((p_{5}-1)a)\label{OC vivo-three-state stab add-bvp11}\\
	&[\mathrm{U}](0) = P_{total}, [\mathrm{I}](0)=[\mathrm{F}](0)= [\mathrm{ISI}](0) = [\mathrm{S}](0) = 0 \\ &\mathbf{p}(T)=(0,0,0,w_{Tol},0).
\end{aligned}
$$

In this case, $p_5$ acts as the key adjoint state.

\subsection{Forms of the Bang-Bang control}
\label{Appendix C3}
We have provided the necessary conditions for the optimal control problem of stabilizing proteins with chemicals. In this section, we will take the in-vitro two-state model as an example to formally demonstrate that the Bang-Bang control has only one switching point. Moreover, before this switching point, the control input $u(t)$ is fixed at $u_{\max}$, while after the switching point, it becomes $0$.

From \eqref{in vitro two-state bvp4}, it is clear to see that the optimal solution is a Bang-Bang control, with at most one switch, and the maximum dosage rate is used at the beginning. Therefore, we only need to show that $p_3$ is strictly monotonically decreasing.  Due to \eqref{in vitro two-state bvp3}, it is equivalent to show \(p_{2}+p_{3}<0\). Considering
$$
\begin{aligned}
&\dot{p_{2}}(t)- \dot{p_{1}}(t) = (k^{+}_{uf}+k^{-}_{uf})(p_{2}-p_{1})+k^{+}_{fs}[\mathrm{S}](p_{2}+p_{3}),\\
&\dot{p_{2}}(t)+ \dot{p_{3}}(t) =k^{-}_{uf}(p_{2}-p_{1})+(k^{-}_{fs}+k^{+}_{fs}[\mathrm{S}]+k^{+}_{fs}[\mathrm{F}])(p_{2}+p_{3}),
\end{aligned}$$
we can introduce
\[A =\left[\begin{array}{ll}
	k^{+}_{uf}+k^{-}_{uf}&k^{+}_{fs}[\mathrm{S}]\\
	k^{-}_{uf}&(k^{-}_{fs}+k^{+}_{fs}[\mathrm{S}]+k^{+}_{fs}[\mathrm{F}])
\end{array}\right]\]
whose two eigenvalues are both greater than zero. Thus, analyzing the ODEs above backward in time, we have a unique steady state, which leads to \(p_{2}+p_{3}<0\). (A more rigorous proof can be done by using the comparison theorem.)
\subsection{Approximation of the optimal switching time}
\label{Appendix C4}
The optimal switching time point equals the time when the key adjoint state $\tilde{p}$ reaches 1, whose explicit form is very complicited. So here we look for an approximate solution, when the terminal time $T$ is sufficiently large to allow each chemical species to approximately reach the equilibrium.
Based on our previous analysis, we have already established that the optimal control takes the following form:
\begin{equation}
	u(t) = \left\{
		\begin{aligned}
			&u_{\max},\quad &0\le t<t_{switch},\\
			&0,\quad &t_{switch}\le t<T,
		\end{aligned}
		\right.
	\end{equation}
where $t_{switch}$ represents the switching time.
Consequently, our objective functional can be transformed into a single-variable optimization problem, that is $$f(t_{switch})\coloneqq J(u).$$

\paragraph{In vitro two-state model}
We introduce the parameter $\psi = w_{Tol}/(K_{uf}+1)>0$. When $\psi \le 1$, the optimal switching time $t_{op} = 0$; when $\psi > 1$, $$t_{op} = \min\{\left[P_{total}-\kappa_{1}+(\psi-2)\sqrt{\kappa_{1}P_{total}/(\psi-1)}\right],T\}.$$ Additionally, when $\psi \gg 1$, we have
$$t_{op} =\min\{\left[P_{total}-\kappa_{1}+\sqrt{\kappa_{1}\psi P_{total}}\right],T\}.$$

Firstly, we establish that when $\psi \leq 1$, $t_{op} = 0$. In this case, $ f(t_{switch}) = w_{Tol}\mathrm{U}(T) + t_{switch}u_{\max} $.
Assuming \( t_{op} > 0 \), we have:
\begin{align*}
	f(t_{op}) =&\frac{w_{Tol}}{2}(\sqrt{\eta_{1}^2+4\kappa_{1} P_{total}}-\eta_{1})/(1+K_{uf})+t_{op}u_{\max} \\
	>& -\psi(\eta_{1}-\kappa_{1})+t_{op}u_{\max}> P_{total}\psi + (1-\psi)t_{op}u_{\max}>f(0).
\end{align*}
This contradicts the definition of $ t_{op} $.

Then, for $ \psi > 1 $, according to the optimization theory, we know that the minimum point of $f $ lies either at the boundary or at a stationary point. Here, we seek for the stationary points of $ f(t) $, which are the points where $ f'(t) = 0 $.
After calculation, we find that the stationary points of $ f $ occur at $$ t =P_{total}-\kappa_{1}+(\psi-2)\sqrt{\kappa_{1}P_{total}/(\psi-1)}.$$
Consequently, we can obtain the optimal switching time point as shown above.

\paragraph{In vitro three-state model}
We introduce the parameter $\psi = w_{Tol}/(\widetilde{K_{uf}}+1)>0$. When $\psi \le 1$, $t_{op} = 0$; when $\psi > 1$, $$t_{op} = \min\{\left[P_{total}-\kappa_{2}+(\psi-2)\sqrt{\kappa_{2}P_{total}/(\psi-1)}\right],T\}.$$ Additionally, when $\psi \gg 1$, the optimal switching time can be approximated as
$$t_{op} =\min\{\left[P_{total}-\kappa_{2}+\sqrt{\kappa_{2}\psi P_{total}}\right],T\}.$$

The analysis here is similar as before.

\paragraph{In vivo two-state model}
Introduce the parameter $\psi = w_{Tol}K_{uf}K_{g}K_{fs}/(1+K_{uf}K_{g}K_{fs})>0.$ In this model, we know that $ f(t_{switch}) = -w_{Tol}\mathrm{FSF}(T) + t_{switch}u_{\max}  $, which implies
\begin{align*}
f(t_{switch}) =& -w_{Tol}(K_{uf}K_{g}+\psi t_{switch}u_{\max}/w_{Tol})+ t_{switch}u_{\max},\\
=&f(0)+(1-\psi)t_{switch}u_{\max}.
\end{align*}
Therefore, when $ \psi \leq 1 $, we have $ t_{op} = 0 $; and when $ \psi > 1 $, we have $ t_{op} = T $.

\paragraph{In vivo three-state model}
Introduce the parameter $\psi =w_{Tol}K_{ui}K_{g}K_{is}/(1+K_{ui}K_{g}K_{is})>0$. In this case, it is directly seen that when $\psi \le 1$, $t_{op} = 0$; when $\psi > 1$, $t_{op} = T.$

\section{Nonlinear strategies for stabilizer addition}
\label{Appendix D}
To assess the impacts of nonlinear stategies for stabilizer addition, we replace the linear relationship by a square-root form, i.e. \[\dot{[\mathrm{S}]}(t) = -k^{+}_{fs}[\mathrm{F}][\mathrm{S}]+k^{-}_{fs}[\mathrm{FS}]+\sqrt{u}.\]
Meanwhile, the objective function and control constraints remain the same as previous settings.
Similar to \ref{Appendix C2}, we can derive the necessary conditions for the corresponding optimal control, that is
$$
\begin{aligned}
	&\dot{[\mathrm{U}]}(t) = -k^{+}_{uf}[\mathrm{U}]+k^{-}_{uf}[\mathrm{F}],\\
	&\begin{aligned}
		\dot{[\mathrm{F}]}(t) & = (k^{+}_{uf}-k^{-}_{fs})[\mathrm{U}]-(k^{-}_{uf}+k^{-}_{fs})[\mathrm{F}]\\&-k^{+}_{fs}[\mathrm{F}][\mathrm{S}]+k^{-}_{fs}P_{total},\end{aligned}\\
	&\dot{[\mathrm{S}]}(t) =-k^{-}_{fs}[\mathrm{U}] -k^{-}_{fs}[\mathrm{F}] -k^{+}_{fs}[\mathrm{F}][\mathrm{S}]+k^{-}_{fs}P_{total} + \sqrt{u},\\
	&\dot{p_{1}}(t) = k^{+}_{uf}p_{1}-(k^{+}_{uf}-k^{-}_{fs})p_{2}+k^{-}_{fs}p_{3},\\
	&\begin{aligned}
		\dot{p_{2}}(t)  =&-k^{-}_{uf}p_{1}+(k^{-}_{uf}+k^{-}_{fs}+k^{+}_{fs}[\mathrm{S}])p_{2}\\&+(k^{-}_{fs}+k^{+}_{fs}[\mathrm{S}])p_{3},
	\end{aligned}\\
	&\dot{p_{3}}(t)  = k^{+}_{fs}[\mathrm{F}]p_{2}+k^{+}_{fs}[\mathrm{F}]p_{3},\\
	&(p_{3}-1)u  = \max_{a\in [0, u_{\max}]}(p_{3}\sqrt{a}-a),\\
	&[\mathrm{U}](0) = P_{total}, [\mathrm{F}](0) = [\mathrm{S}](0) = 0, \mathbf{p}(T)=(-w_{Tol},0,0).
\end{aligned}
$$
The solution for the optimal strategy can be derived as
\begin{equation}
	u(t) = \left\{
	\begin{aligned}
		&u_{\max},\quad &p_{3} \ge 2\sqrt{u_{max}},\\
		&p_{3}^2/4,\quad &0<p_{3}<2\sqrt{u_{max}},\\
		&0,\quad &p_{3} \le 0.
	\end{aligned}
	\right.
\end{equation}

\section{Free terminal time}\label{Appendix E}
Once the terminal time is not fixed, we modify the objective function as
\begin{equation}
	J[u(\cdot),t_{free}] = \sigma t_{free}+\int_{0}^{t_{free}}u(\tau)d \tau,
\end{equation}
where the free terminal time \( t_{free} \) is a variable, and $\sigma$ is the weight associated with the terminal time. The terminal constraint on the amount of folded proteins is given by
\begin{equation}
	[\mathrm{FSF}](t_{free}) = [\mathrm{FSF}]_{tar},
\end{equation}
where \( [\mathrm{FSF}]_{tar} \) represents the total amount of folded proteins at target. The controlled kinetic equations and control constraints remain the same as the settings in Section \ref{subsed2.3}. Here, the Pontryagin Maximum Principle is utilized for the free terminal time.  Similar to results in \ref{Appendix C2}, we can derive the necessary conditions for the corresponding optimal control,
$$
\begin{aligned}
	&\dot{[\mathrm{U}]}(t) = -k^{+}_{uf}[\mathrm{U}]+k^{-}_{uf}[\mathrm{F}], \\
	&\begin{aligned}
		\dot{[\mathrm{F}]}(t) & = (k^{+}_{uf}-k^{-}_{fs})[\mathrm{U}]-(k^{-}_{uf}+k^{-}_{fs})[\mathrm{F}]\\&-k^{+}_{fs}[\mathrm{F}][\mathrm{S}]+k^{-}_{fs}P_{total},\end{aligned}\\
	&\dot{[\mathrm{S}]}(t) =-k^{-}_{fs}[\mathrm{U}] -k^{-}_{fs}[\mathrm{F}] -k^{+}_{fs}[\mathrm{F}][\mathrm{S}]+k^{-}_{fs}P_{total} + u,\\
	&\dot{p_{1}}(t) = k^{+}_{uf}p_{1}-(k^{+}_{uf}-k^{-}_{fs})p_{2}+k^{-}_{fs}p_{3},\\
	&\begin{aligned}
		\dot{p_{2}}(t)  =&-k^{-}_{uf}p_{1}+(k^{-}_{uf}+k^{-}_{fs}+k^{+}_{fs}[\mathrm{S}])p_{2}\\&+(k^{-}_{fs}+k^{+}_{fs}[\mathrm{S}])p_{3},
	\end{aligned}\\
	&\dot{p_{3}}(t)  = k^{+}_{fs}[\mathrm{F}]p_{2}+k^{+}_{fs}[\mathrm{F}]p_{3},\\
	&(p_{3}-1)u  = \max_{a\in [0, u_{\max}]}((p_{3}-1)a),\\
	&[\mathrm{U}](0) = P_{total}, [\mathrm{F}](0) = [\mathrm{S}](0) = 0, [\mathrm{U}](t_{free}) = P_{total}- \mathrm{FSF}_{tar}\\
	&(p_{2}(t_{free}),p_{3}(t_{free}))=(0,0),\\
	&H([\mathrm{U}](t),[\mathrm{F}](t),[\mathrm{S}](t),p_{1}(t),p_{2}(t),p_{3}(t),u(t))|_{t=t_{free}}=0,
\end{aligned}
$$
where the Hamiltonian reads
$$
\begin{aligned}
	&H([\mathrm{U}],[\mathrm{F}],[\mathrm{S}],p_{1},p_{2},p_{3},u) = \\
	&(-k^{+}_{uf}[\mathrm{U}]+k^{-}_{uf}[\mathrm{F}])p_{1}\\
	&+\left((k^{+}_{uf}-k^{-}_{fs})[\mathrm{U}]-(k^{-}_{uf}+k^{-}_{fs})[\mathrm{F}]-k^{+}_{fs}[\mathrm{F}][\mathrm{S}]+k^{-}_{fs}P_{total}\right)p_{2}\\
	&+\left(-k^{-}_{fs}[\mathrm{U}] -k^{-}_{fs}[\mathrm{F}] -k^{+}_{fs}[\mathrm{F}][\mathrm{S}]+k^{-}_{fs}P_{total} + u\right)p_{3}-u-\sigma.
\end{aligned}
$$
This is the most dramatic difference compared to the fixed terminal time problem.

\section{Optimal controls in the presence of protein aggregation}
\label{Appendix F}
With respect to the in-vitro three-state model for protein folding kineitcs and the sequential elongation model for protein aggregation, the full kinetic equations can be derived, i.e. 
\begin{equation}\label{Appendix three-state-cascade-reaction}
	\begin{aligned}
		\dot{[\mathrm{U}]}(t)  =& -k^{+}_{ui}[\mathrm{U}]+k^{-}_{ui}[\mathrm{I}],\\
		\dot{[\mathrm{F}]}(t)  =& k^{+}_{if}[\mathrm{I}]-k^{-}_{if}[\mathrm{F}],\\
		\dot{[\mathrm{IS}]}(t)  =& k^{+}_{is}[\mathrm{I}][\mathrm{S}]-k^{-}_{is}[\mathrm{IS}],\\
		\dot{[\mathrm{S}]}(t)  =& -k^{+}_{is}[\mathrm{I}][\mathrm{S}]+k^{-}_{is}[\mathrm{IS}]+u.\\
		\dot{[\mathrm{I}]}(t) =&k^{+}_{ui}[\mathrm{U}]-k^{-}_{ui}[\mathrm{I}] - k^{+}_{if}[\mathrm{I}]+k^{-}_{if}[\mathrm{F}]-k^{+}_{is}[\mathrm{I}][\mathrm{S}]\\
		&+k^{-}_{is}[\mathrm{IS}]+2k^{-}[\mathrm{I_{2}}]-2k^{+}[\mathrm{I}]^2+\sum_{j=3}^{n}\left(k^{-}[\mathrm{I_{j}}]-k^{+}[\mathrm{I_{j-1}}][\mathrm{I}]\right),\\
		\dot{[\mathrm{I_{j}}]}(t)=&k^{+}([\mathrm{I_{j-1}}]-[\mathrm{I_{j}}])[\mathrm{I}]+k^{-}([\mathrm{I_{j+1}}]-[\mathrm{I_{j}}]),\quad j = 2,\ldots,n-1, \\
		\dot{[\mathrm{I_{n}}]}(t)=&k^{+}[\mathrm{I_{n-1}}][\mathrm{I}]-k^{-}[\mathrm{I_{n}}].
	\end{aligned}
\end{equation}

The objective function and the constraints remain unchanged, as given by Eq.\eqref{Appendix objective functional three-state} and Eq.\eqref{Appendix constraints}, respectively. Similar to \ref{Appendix C2}, we can derive the necessary conditions for the corresponding optimal control, that is
$$
\begin{aligned}
	&\dot{[\mathrm{U}]}(t)  = -k^{+}_{ui}[\mathrm{U}]+k^{-}_{ui}[\mathrm{I}],\\
&\begin{aligned}
	\dot{[\mathrm{I}]}(t) & = (k^{+}_{ui}-k^{-}_{if})[\mathrm{U}]-(k^{-}_{ui}+k^{+}_{if}+k^{-}_{is})[\mathrm{I}]\\
		&+(k^{-}_{is}-k^{-}_{if})[\mathrm{ISI}]-k^{+}_{is}[\mathrm{I}][\mathrm{S}]+k^{-}_{if}P_{total}\\
		&+\sum_{j=2}^{n}\left(k^{-}[\mathrm{I_{j}}]-k^{+}[\mathrm{I_{j-1}}][\mathrm{I}]\right)+k^{-}[\mathrm{I_{2}}]-k^{+}[\mathrm{I}]^2,
	\end{aligned}
	\\
	&\dot{[\mathrm{ISI}]}(t) = (k^{+}_{ui}-k^{-}_{if})[\mathrm{U}]-(k^{-}_{ui}+k^{+}_{if})[\mathrm{I}]-k^{-}_{if}[\mathrm{ISI}]+k^{-}_{if}P_{total},\\
	&\dot{[\mathrm{S}]}(t) = -k^{-}_{is}[\mathrm{I}]+k^{-}_{is}[\mathrm{ISI}]-k^{+}_{is}[\mathrm{I}][\mathrm{S}]+u,\\
	&\dot{[\mathrm{I_{j}}]}(t)=k^{+}([\mathrm{I_{j-1}}]-[\mathrm{I_{j}}])[\mathrm{I}]+k^{-}([\mathrm{I_{j+1}}]-[\mathrm{I_{j}}]),\quad j = 2,\ldots,n-1, \\
	&\dot{[\mathrm{I_{n}}]}(t)=k^{+}[\mathrm{I_{n-1}}][\mathrm{I}]-k^{-}[\mathrm{I_{n}}],\\
	&\dot{p_{1}}(t) = k^{+}_{ui}p_{1}-(k^{+}_{ui}-k^{-}_{if})(p_{2}+p_{3}),\\
	&\dot{p_{2}}(t) =-k^{-}_{ui}p_{1}+(k^{-}_{ui}+k^{+}_{if})(p_{2}+p_{3})+(k^{-}_{is}+k^{+}_{is}[\mathrm{S}])(p_{2}+p_{4}),\\
	&+k^{+}\left[4[\mathrm{I}]p_{2}- 2[\mathrm{I}]q_{2}+\sum_{j=2}^{n-1}[\mathrm{I_{j}}]\left(p_{2}+q_{j}-q_{j+1}\right)\right] \\
	&\dot{p_{3}}(t) =k^{-}_{if}(p_{2}+p_{3})-k^{-}_{is}(p_{2}+p_{4}),\\
	&\dot{p_{4}}(t) = k^{+}_{is}[\mathrm{I}](p_{2}+p_{4}),\\
	&\dot{q_{2}}(t) = -k^{-}(2p_{2}-q_{2})+k^{+}[\mathrm{I}](p_{2}+q_{2}-q_{3}),\\
	&\dot{q_{j}}(t) = -k^{-}(p_{2}+q_{j-1}-q_{j})+k^{+}[\mathrm{I}](p_{2}+q_{j}-q_{j+1}), j=3,\ldots, n-1,\\
	&\dot{q_{n}}(t) = -k^{-}(p_{2}+q_{n-1}-q_{n}),\\
\end{aligned}
$$
$$
\begin{aligned}
	&(p_{4}-1)u = \max_{a\in [0, u_{\max}]}((p_{4}-1)a),\\
	&[\mathrm{U}](0) =P_{total}, [\mathrm{I}](0) = [\mathrm{ISI}](0) = [\mathrm{S}](0) = 0, \\
	& \mathbf{p}(T)=(0,0,w_{Tol},0),q_{j}(T)=0,j=2,\ldots,n.
\end{aligned}
$$
Accordingly, the solution for the optimal strategy takes a bang-bang form,
\begin{equation}
	u(t) = \left\{
	\begin{aligned}
		&u_{\max},\quad &p_{4} \ge 1,\\
		&0,\quad &p_{4} < 1.
	\end{aligned}
	\right.
\end{equation}








\end{document}